\documentclass[intlimits,twoside,a4paper]{article}

\usepackage{amsmath,amssymb}
\usepackage{graphicx}

\usepackage[T2A]{fontenc}
\usepackage[cp1251]{inputenc}

\usepackage[eqsecnum]{cmpj2}
\issue{2015}{18}{3}{33001}
\doinumber{10.5488/CMP.18.33001}

\title[On relaxation phenomena in a two-component plasma]
{On relaxation phenomena in a two-component plasma}

\author[V.N.~Gorev, A.I.~Sokolovsky, Z.Yu.~Chelbaevsky]{V.N.~Gorev, A.I.~Sokolovsky, Z.Yu.~Chelbaevsky}
\address{Oles Honchar Dnipropetrovsk National University, 72 Gagarin Ave., 49010 Dnipropetrovsk, Ukraine}

\date{Received September 30, 2014, in final form March 26, 2015}
\authorcopyright{V.N.~Gorev, A.I.~Sokolovsky, Z.Yu.~Chelbaevsky, 2015}

\begin{document}

\maketitle

\begin{abstract}
The relaxation of temperatures and velocities of the components of a quasi-equilibrium two-component homogeneous completely ionized plasma is investigated on the basis of a generalization of the Chapman-Enskog method applied to the Landau kinetic equation. The generalization is based on the functional hypothesis in order to account for the presence of kinetic modes of the system. In the approximation of a small difference of the component temperatures and velocities, it is shown that relaxation really exists (the relaxation rates are positive). The proof is based on the arguments that are valid for an arbitrary two-component system. The equations describing the temperature and velocity kinetic modes of the system are investigated in a perturbation theory in the square root of the small electron-to-ion mass ratio. The equations of each order of this perturbation theory are solved with the help of the Sonine polynomial expansion. Corrections to the known Landau results related to the distribution functions of the plasma and relaxation rates are obtained. The hydrodynamic theory based on these results should take into account a violation of local equilibrium in the presence of relaxation processes.
\keywords distribution function, Chapman-Enskog method, relaxation rate, Sonine polynomials, kinetic equation, kinetic modes, fully ionized plasma
\pacs 02.30.Rz, 05.20.Dd, 52.25.Dg
\end{abstract}

\section{Introduction}

In his known paper \cite{Land} Landau obtained a kinetic equation
for a two-component fully ionized elec\-tron-ion plasma. This equation is widely used for investigation of plasma kinetics (see, for example, \cite{Landau_81, Silin,  Brantov1, Brantov2,  BobPotSak}). Of course, it describes the situation in the plasma approximately. The Landau equation takes into account only the short-range part of the Coulomb interaction because the Coulomb potential is artificially cut in the collision integral at the Debye radius. This can be done exactly using the Balescu-Lenard equation. In the Landau collision integral, the Coulomb potential is also cut at small distances where this potential is big and the situation needs special attention. This was done in an exact consideration by Rukhadze and Silin \cite{SilRukh}. A comparison of the mentioned theories shows that the Landau kinetic equation describes the effects of the short-range part of the Coulomb interaction in plasma with a logarithmic accuracy (see the discussion of the mentioned results in \cite{Landau_81}). The long-range part of the Coulomb interaction can be taken into account by  the Vlasov term which describes the one-particle effects of a self-consistent field \cite{Landau_81}. In homogeneous states, there is no self-consistent field, and one can investigate these states only on the basis of the Landau kinetic equation.

On the basis of his equation \cite{Land} Landau investigated the case in which the components are spatially homogeneous equilibrium subsystems with different temperatures $T_a(t)$ ($a=\textrm{e,i}$) and the temperature relaxation is observed. This phenomenon is of great interest because of its fundamental importance for applications in plasma theory and condensed matter physics in general. Temperature and velocity relaxation in two-component systems is observed in many systems and breaks local equilibrium in them. Even for spatially uniform states, the problem of calculating the distribution function in the presence of relaxation is considered to be a complicated one  \cite{BobPotSak} because of the lack of a small parameter. Among the existing applications, it is worth to mention the  plasma hydrodynamics with account of the relaxation processes (two fluid hydrodynamics) \cite{Braginsky1,Braginsky2}, electron-phonon hydrodynamic phenomena in metals and semiconductors \cite{TTM}, magnon-phonon hydrodynamic phenomena in ferromagnetic materials  \cite{AkhBarPel}, relaxation of the hot spot \cite{Brantov2013}, etc.

In the present paper, the relaxation of temperatures $T_a(t)$ and velocities $\upsilon_{an}(t)$ of the plasma components
\begin{equation}\label{Relaxation}
T_a(t)\xrightarrow[t\gg\tau_T]{}T,\qquad \upsilon_{an}(t)\xrightarrow[t\gg\tau_u]{}\upsilon_n	
\end{equation}
 are investigated (on the basis of the ideas of paper \cite{Land} the velocity relaxation was studied, for example, in \cite{AlBogdRukh}). In \eqref{Relaxation}, $T, \upsilon_n$ are the equilibrium temperature and velocity of the plasma; $\tau_T, \tau_u$ are the relaxation times.  Our consideration is based on the Chapman-Enskog method {\it generalized to account for the relaxation phenomena} (we use the term ``relaxation phenomena'' in the narrow sense of the word as nonequilibrium processes that can be observed in spatially uniform states). Such a theory should describe kinetic modes of the system. In the recent statistical mechanics, the problem of investigating the effect of kinetic modes on the behavior of nonequilibrium systems is considered to be very important (see, for instance, review \cite{Mr}). The generalization is based on the idea of the Bogoliubov functional hypothesis (see, for example, \cite{AkhPel,Colangeli}), which describes the structure of the nonequilibrium distribution function at the times under consideration
\begin{equation}\label{FunctHyp}
f_{ap}(t)\xrightarrow[t\gg\tau_0]{}f_{ap}\left( T_\textrm{e}(t,f_0), \upsilon_\textrm{e}(t,f_0)\right),\qquad \left[\tau_0\ll\tau_T,\tau_u, \quad
f_{ap0}\equiv f_{ap}(t=0), \quad {f_0} \equiv \left\{ {{f_{ap0}}} \right\}\right].
\end{equation}
The function $f_{ap}(T_\textrm{e}(t,f_0), \upsilon_\textrm{e}(t,f_0))$ contains asymptotic values of the parameters
\begin{equation}\label{RDP}
T_\textrm{e}(t)\xrightarrow[t\gg\tau_0]{}T_\textrm{e}(t,f_0), \qquad \upsilon_\textrm{e}(t)\xrightarrow[t\gg\tau_0]{}\upsilon_\textrm{e}(t,f_0).
\end{equation}

So, at times $t\gg\tau_0$, the distribution function depends on the time and the initial state of the system $f_{ap0}$ only through parameters that describe the state of the system (the reduced description parameters). Here, $\tau_0$ is some characteristic time which is chosen to precede the end of relaxation processes. In fact, only the set of reduced description parameters depends on $\tau_0$, but the value of $\tau_0$ depends on the initial state of the system $f_{ap0}$. Arrows in the functional hypothesis \eqref{FunctHyp} and definitions \eqref{RDP} show that their right-hand sides are {\it a result of the natural evolution of the system}. The function $f_{ap}(T_\textrm{e}(t,f_0), \upsilon_\textrm{e}(t,f_0))$ {\it is the asymptotic limit of the distribution function $f_{ap}(t),$ and it exactly satisfies the kinetic equation}. The asymptotic distribution $f_{ap}(T_\textrm{e}, \upsilon_\textrm{e})$ does not depend on the initial state of the system $f_{ap0}$. These statements are the basic ideas concerning the functional hypothesis applied by us for a generalization of the Chapman-Enskog method. The reduced description parameters in \eqref{FunctHyp} do not include the ion temperature and velocity due to the energy and momentum conservation laws in the spatially uniform states under consideration. They are functions of the electron temperature and velocity (in what follows we do not show the dependence of the reduced description parameters on the initial distribution function $f_{ap0}$).

The Landau approximation \cite{Land} (and the corresponding approximation of \cite{AlBogdRukh}) can be written in the form
\begin{equation}\label{LandauDF}
f_{ap}(T_\textrm{e}, \upsilon_\textrm{e})=w_{a,p-m_a\upsilon_a}(T_a), \qquad w_{ap}(T_a)\equiv\frac{n_a}{(2\pi m_a T_a)^{3/2}}\exp\left\{-\frac{\varepsilon_{ap}}{T_a}\right\}
\end{equation}
($\varepsilon_{ap}\equiv p_a^2/2m_a$). The use of the Maxwell distribution for a system consisting of interacting equilibrium subsystems (quasi-equilibrium state) is quite attractive from the physical point of view. The corresponding local distribution gives the local equilibrium approximation for the description of spatially nonuniform states. This simple idea by Landau is a basis of many investigations. For example, plasma hydrodynamics was investigated in \cite{Braginsky1,Braginsky2} on the basis of the Landau approximation. The local equilibrium approximation is a basis of investigations in the book \cite{Zhdanov} devoted to transport processes in a multicomponent plasma.

In the present paper, the problem of correction of the assumption \eqref{LandauDF} is considered as a very important one, and {\it the distribution function} $f_{ap}(T_\textrm{e}, \upsilon_\textrm{e})$ {\it is calculated in a perturbation theory in a small difference of the component temperatures and velocities} (let the corresponding small parameter be $\lambda $). In terms of the theory of hydrodynamic states, this means that in the plasma in the presence of relaxation, the local equilibrium is violated. Note, that in spatially inhomogeneous states, relaxation in a two-component system was studied in \cite{BatsMrRudTok}. However, the authors of \cite{BatsMrRudTok} did not obtain these results because they described the system by the energy densities of the components. {\it The idea of considering the relaxation processes in the system at their end was proposed in our paper} \cite{GorSokChelb} and presented at the conferences QEDSP 2011 and MECO 38.
It is worth noting that this idea can be also used in the Grad method.
As is known \cite{ExtTh}, the disadvantage of this method is the absence of a small
parameter. The above-mentioned parameter $\lambda$ can be selected for a small parameter in the
Grad method. Further development of this approach was presented by us in \cite{NASMath}.

The final results of the present paper are given in an additional perturbation theory in the small mass ratio $\sigma\equiv \sqrt{m_\textrm{e}/m_\textrm{i}}$,  and the integral equations of the theory in each order in $\sigma$ are solved using the method of truncated Sonine polynomial expansion.

The paper is organized as follows. In section~\ref{BEq}, the basic definitions and equations of the theory are presented. In section~\ref{GenCE}, a generalized Chapman-Enskog method is developed and integral equations for the distribution functions are obtained. In section~\ref{AppSol}, these equations are solved in a $\sigma$ perturbation theory with the help of the truncated Sonine polynomial expansion method.

\section{Basic equations of the theory \label{BEq}}

A two-component fully ionized electron-ion plasma can be described by the Landau kinetic equation \cite{Land}. In the considered case of spatially uniform states, the component distribution function $f_{ap}$ satisfies the equation
\begin{equation}
\label{LandauKE}
\frac{{\partial {f_{ap}(t)}}}{{\partial t}} = I_{ap}(f(t))
\end{equation}
with  the standard expression for the collision integral $I_{ap}(f)$
\[
I_{ap}(f) = 2\pi e_a^2 L \sum_b  e_b^2 \frac{\partial }{\partial p_n}\int \rd^3p'
D_{nl}\left(\frac{p}{m_a}-\frac{p'}{m_b}\right)\left(\frac{\partial f_{ap}}{\partial p_l}f_{bp'} - f_{ap}\frac{\partial f_{bp'}}{\partial p'_l} \right ),
\]
\begin{equation}\label{LandauCI}
D_{nl}(u) \equiv (u^2\delta_{nl}-u_nu_l)/|u|^3\,,
\end{equation}
where $L$ is the Coulomb logarithm (the subscripts $a,b,c,\ldots= \textrm{e,\,i}$ denote the electron and ion components). The quantities $m_a$, $e_a$  are the particle masses and charges ${e_\textrm{e}} =  - e$, ${e_\textrm{i}} = z e$, where $e$  is the elementary electric charge and $z$ is the ion charge number.

The particle number density $n_a$, temperature $T_a$ and velocity $\upsilon_{an}$ of the components are defined using the standard formulas \cite{Silin,AkhPel}
\begin{equation} \label{def_ntv}
{n_a} = \int \rd^3p {{f_{ap}}}\,, \qquad
{\pi _{an}} = {m_a}{n_a}{\upsilon _{an}} = \int \rd^3p {{f_{ap}}{p_n}}\,,
\qquad{\varepsilon _a} = \frac{3}{2}{n_a}{T_a} + \frac{1}{2}m_a n_a{\upsilon_a}^2 = \int \rd^3p {{f_{ap}}\varepsilon_{ap}}\,,
\end{equation}
where  $\pi_{an}$ and $\varepsilon _a$  are the momentum and energy densities of the components. In this paper, the temperature is measured in energy units.

Let us introduce the quantities $\upsilon_n $ and $T$  as:
\begin{equation}
\label{def_tv}
\pi _n = \sum_a \pi _{an} = \upsilon _n\sum_a m_a n_a\,, \qquad  \varepsilon  = \sum_a \varepsilon _a = \frac{3}{2}T\sum_a n_a  + \frac{1}{2}\upsilon^2\sum_a m_a n_a\,,
\end{equation}
where $\pi _n$, $\varepsilon $ are the total momentum and the total energy densities, respectively. These quantities do not depend on time because of the relations
\begin{equation} \label{Sources}
\quad \frac{\partial \pi_{an}(t)}{\partial t}=R_{an}(f(t)), \qquad \frac{\partial \varepsilon_a(t)}{\partial t}=Q_a(f(t))\,,
\end{equation}
where the functions
\begin{equation} \label{def_Sources}
R_{an}(f)\equiv\int \rd^3p p_nI_{ap}(f), \qquad Q_a(f)\equiv\int \rd^3p \varepsilon_{ap} I_{ap}(f) 	
\end{equation}
obey the properties
\begin{equation} \label{TotalSources}
\sum_a R_{an}(f)=0, \qquad \sum_a Q_a(f)=0.	
\end{equation}
The particle densities $n_a$ do not depend on time because
\begin{equation} \label{CI_property}
\int \rd^3p I_{ap}(f)=0.	
\end{equation}
According to the definition \eqref{def_ntv}, \eqref{def_tv}, the quantities $\upsilon_n $, $T$ are equal to the equilibrium temperature and velocity of the system and can be considered as the given parameters. According to the concept of the Galilean invariance, we consider the problem under study in the reference frame where $\upsilon_n=0$.

Let us introduce the deviations $\tau, u_n$ of the electron subsystem temperature $T_\textrm{e}$ and velocity $\upsilon_{\textrm{e}n}$ from their equilibrium values
\begin{equation} \label{def_Deviations}
T_\textrm{e}-T= \tau, \qquad \upsilon_{\textrm{e}n} = u_n
\end{equation}
which should be considered as parameters describing the relaxation in the system because the deviations of the ion temperature $T_\textrm{i}$ and velocity $\upsilon_{\textrm{i}n}$ are also expressed in terms of $\tau$, $u_n$
\begin{equation} \label{IonDeviations}
T_\textrm{i}-T =  -z\tau  - \frac{1}{3}m_\textrm{e} z u^2( 1 + z\sigma^2),\qquad \upsilon_{\textrm{i}n} =  - z\sigma^2 u_n\,, \qquad \left[\sigma=(m_\textrm{e}/m_\textrm{i})^{1/2}\right].
\end{equation}
Here, it was taken into account that in spatially uniform states, the charge neutrality condition $n_\textrm{i}\equiv n$, $n_\textrm{e} = zn$ is satisfied.  Expressions \eqref{IonDeviations} justify the functional hypothesis  in the form \eqref{FunctHyp} that contains only electron variables.

\section{Generalization of the Chapman-Enskog method \label{GenCE}}

The generalization of the Chapman-Enskog method presented here is based on the functional hypothesis \eqref{FunctHyp}, which can be written in the form
\begin{equation} \label{FunctHyp_n}
f_{ap}(t)\xrightarrow[t\gg\tau_0]{}f_{ap}(\tau(t), u(t))
\end{equation}
suitable for our consideration of the relaxation processes at their end. Then, substitution of \eqref{def_ntv} into \eqref{Sources} with account for \eqref{def_Deviations} leads at $t\gg \tau_0$ to the closed-form time equations for the parameters $\tau$, $u_n$
\begin{equation} \label{TimeEquations}
\frac{{\partial {u_n}(t)}}{{\partial t}} = \frac{1}{ n z  m_\textrm{e}}R_{\textrm{e}n}(f(\tau(t),u(t))),\qquad \frac{\partial \tau(t) }{\partial t} = \frac{2}{3nz}Q_\textrm{e}(f(\tau(t),u(t))).	
\end{equation}
According to the basic idea of the reduced description method, the distribution function $f_{ap}(\tau(t),u(t))$
exactly satisfies the kinetic equation \eqref{LandauKE} for times $t \gg {\tau _0}$. This leads to the following integro-differential equation for the function $f_{ap}(\tau,u)$
\begin{equation} \label{KEfg}
\frac{\partial {f_{ap}}(\tau,u)}{\partial {u_n}}\frac{1}{ n z m_\textrm{e}}R_{\textrm{e}n}(f(\tau,u)) + \frac{\partial f_{ap}(\tau,u)}{\partial \tau }\frac{2}{{3 n z}}Q_\textrm{e}(f(\tau,u)) = I_{ap}(f(\tau,u)).
\end{equation}

One should add to this equation the definition of the parameters $u_n$ and $\tau $ given in terms of the distribution function $f_{ap}(\tau, u)$
\[
\int \rd^3 p f_{ap}(\tau,u)=n_a\,,\qquad \int \rd^3 p p_nf_{ap}(\tau,u)=m_a n_a \upsilon_{an}(\tau,u)\,,
\]
\begin{equation} \label{AdditionalConditions}
\int \rd^3 p \varepsilon_{ap} f_{ap}(\tau,u)=\frac{3}{2}n_a T_a(\tau,u)+\frac{1}{2}m_a n_a\upsilon_{a}(\tau,u)^2\,,
\end{equation}
where the component temperatures and velocities $T_a(\tau,u)$, $\upsilon_{an}(\tau,u)$ as functions of $\tau,u$ are defined by the formulas
\begin{equation} \label{vT_C}
{\upsilon_{\textrm{e}n}}(\tau,u) = u_n\,, \qquad  \upsilon_{\textrm{i}n}(\tau,u) =  - z\sigma^2 u_n\,, \qquad T_\textrm{e}(\tau,u) = T + \tau\,, \qquad   T_\textrm{i}(\tau,u) = T -z\tau  - \frac{1}{3}m_\textrm{e} z u^2\left(1 + z\sigma^2\right).	
\end{equation}

In the present paper, relaxation processes in the system are investigated at their end. The corresponding small parameter $\lambda$ can be introduced by estimates
\begin{equation} \label{Lambda}
\frac{\tau}{T}\sim \lambda, \qquad \frac{u_n}{\sqrt{T/m_\textrm{e}}}\sim \lambda, \qquad (\lambda \ll 1).
\end{equation}
The solution of equation \eqref{KEfg} with additional conditions \eqref{AdditionalConditions} is found in the form of a series in $\lambda$
\begin{equation}\label{F_series}
{f_{ap}}(\tau,u) = f_{ap}^{(0)} + f_{ap}^{(1)} + O(\lambda^2).
\end{equation}
The further calculation needs only the assumption that $u_n$, $\tau \sim\lambda$ $(\lambda \ll 1)$. In fact, estimates \eqref{Lambda} follow from the requirement $|f_{ap}^{(1)}|\ll f_{ap}^{(0)}$ and the expression for $f_{ap}^{(1)}$ obtained below.

In the main approximation, equations \eqref{KEfg}, \eqref{AdditionalConditions} give the Maxwell distribution with the equilibrium temperature
\begin{equation} \label{Maxwell}
f_{ap}^{(0)} = w_{ap}\,, \qquad [w_{ap}\equiv w_{ap}(T)].
\end{equation}
This is true because the distribution $w_{ap}$ meets additional conditions \eqref{AdditionalConditions}
\begin{equation} \label{Maxwell_AC}
\langle 1 \rangle_a=n_a\,,\qquad \langle p_n \rangle_a=0\,, \qquad
\langle \varepsilon_{ap}\rangle_a=\frac{3}{2}n_a T\,,
\end{equation}
it does not depend on $\tau$, $u_n$ and is an equilibrium distribution $I_{ap}(w)= 0$ (hereafter for an arbitrary function $g_p$, the notation
\begin{equation}\label{av}
\langle g_p\rangle_a = \int \rd^3p w_{ap}g_p
\end{equation}
is used).

In view of rotational invariance, the solution of equation \eqref{KEfg} in the first order in $\lambda$ has the structure
\begin{equation} \label{F1}
f_{ap}^{(1)}(\tau,u) = w_{ap}[A_a(\beta \varepsilon_{ap})\tau  + B_a(\beta \varepsilon_{ap})p_{n} u_n], \qquad (\beta\equiv T^{-1})
\end{equation}
where $A_a(x)$, $B_a(x)$ are some scalar functions. Note that in the Landau approximation \eqref{LandauDF}, these functions are given by the relations
\begin{equation}\label{AB_Land}
A_\textrm{e}^\textrm{L}(x)=\beta(x-3/2),\quad A_\textrm{i}^\textrm{L}(x)=-z\beta(x-3/2);\qquad B_\textrm{e}^\textrm{L}(x)=\beta,\quad B_\textrm{i}^\textrm{L}(x)=-z\beta\sigma^2.
\end{equation}
Substitution of \eqref{F1} into \eqref{def_Sources} gives the right-hand sides $R_{\textrm{e}n}(f(\tau,u))$, $Q_\textrm{e}(f(\tau,u))$ of equations \eqref{TimeEquations} in the first order in $\lambda$
\begin{equation}
\label{g_30}
R_{\textrm{e}n}^{(1)}=-m_\textrm{e} n z\lambda_u u_n\,, \qquad Q_\textrm{e}^{(1)}=-\frac{3}{2}n z\lambda_T \tau 	\,,
\end{equation}
where the notations
\begin{equation} \label{RelaxationRates}
\lambda_u=\frac{1}{3 m_\textrm{e} n z}\sum_a\left\{p_{n},B_a(\beta\varepsilon_{ap}) p_{n}\right\}_{\textrm{e}a}\,, \qquad
\lambda_T=\frac{2}{3 n z}\sum_a \left\{\varepsilon_{\textrm{e}p},A_a(\beta\varepsilon_{ap})\right\}_{\textrm{e}a}	
\end{equation}
are introduced (these formulas are written in terms of the integral brackets $\{g_p,h_p\}_{ab}$ defined by \eqref{A_IntegralBracket} in the appendix). Substitution of expressions \eqref{g_30} into \eqref{TimeEquations} gives the evolution equations for the parameters $\tau$, $u$
\begin{equation} \label{TimeEquationsLambda}
\frac{\partial \tau}{\partial t}=-\lambda_T\tau+O(\lambda^2), \qquad
\frac{\partial u_n}{\partial t}=-\lambda_u u_n+O(\lambda^2)
\end{equation}
which describe their relaxation in the main approximation. Thus, the quantities $\lambda_T$, $\lambda_u$ are the relaxation rates, and $\tau_T\equiv\lambda_T^{-1}$, $\tau_u\equiv \lambda_u^{-1}$ are the corresponding relaxation times for the temperature and velocity.

The contribution $f_{ap}^{(1)}$ \eqref{F_series}, according to \eqref{KEfg} and \eqref{RelaxationRates}, satisfies the equation
\begin{equation} \label{EquationForF1}
-\lambda_u u_n \frac{\partial f_{ap}^{(1)}}{\partial u_n} - \lambda_T \tau \frac{\partial f_{ap}^{(1)}}{\partial \tau} = \sum_b \int \rd^3 p' M_{ab} (p,p') f_{bp'}^{(1)}\,.
\end{equation}
The kernel $M_{ab}(p,p')$ in this equation is defined by the linearized Landau collision integral
\eqref{LandauCI}
\begin{equation} \label{LandauCI1}
I_{ap}(w+\delta f)=\sum_b\int \rd^3 p' M_{ab}(p,p')\delta f_{bp'}+O((\delta f)^2).	
\end{equation}
Both sides of equation \eqref{EquationForF1} contain the unknown function $f_{ap}^{(1)}$. So, {\it our choice of the small parameter $\lambda$ of the theory leads to a generalization of the Chapman-Enskog method too}. Note that the left-hand sides of similar equations for the standard hydrodynamic state which is investigated on the basis of the Chapman-Enskog method contain only the known functions.

Substitution of expression \eqref{F1} into this equation leads to the integral equations for the functions $A_a(\beta\varepsilon_{ap})$ and $B_a(\beta\varepsilon_{ap})$:
\begin{equation} \label{AB_equations}
\lambda _T{A_a}(\beta\varepsilon_{ap}) = \sum_b {\hat K}_{ab}A_b(\beta \varepsilon_{bp}), \qquad
\lambda _u B_a(\beta\varepsilon_{ap})p_n = \sum_b {\hat K}_{ab}B_b(\beta\varepsilon_{bp})p_n\,.	
\end{equation}
Here, $\hat K_{ab}$ is the linearized collision operator defined in the appendix by formulas \eqref{A_Kab_def}.
Additional conditions to equations \eqref{AB_equations} follow from \eqref{AdditionalConditions} and \eqref{F1}, and they can be written in the form
\begin{equation} \label{AddCond}
\langle A_a(\beta\varepsilon_{ap}) \rangle_a=0, \qquad \langle A_a(\beta\varepsilon_{ap})\varepsilon_{ap}\rangle_a=\frac{3}{2}n z(\delta_{a\textrm{e}}-\delta_{a\textrm{i}}), \qquad
\langle B_a(\beta\varepsilon_{ap})\varepsilon_{ap}\rangle_a=\frac{3}{2}n z(\delta_{a\textrm{e}}-\sigma^2\delta_{a\textrm{i}}).
\end{equation}
Equations \eqref{AB_equations}, \eqref{AddCond} are the main equations of the developed theory that will be analyzed in the next part of this paper.

Integral equations \eqref{AB_equations} show that $A_a(\beta \varepsilon _{ap})$, $B_a(\beta\varepsilon_{ap})p_n$ are eigenfunctions and $\lambda_T$, $\lambda_u$ are the corresponding eigenvalues for the linearized collision operator of the kinetic equation under consideration. They describe the temperature and velocity kinetic modes of the system. It is important to emphasize that  formulas \eqref{RelaxationRates} are consequences of integral equations \eqref{AB_equations} and the additional conditions \eqref{AddCond}. Therefore, {\it equations \eqref{AB_equations} with additional conditions \eqref{AddCond} can be solved without taking into account the expressions} \eqref{RelaxationRates}. However, it may be useful to simplify the calculation.

The quantities $\lambda_T$, $\lambda_u$ are positive due to the identities
\begin{eqnarray}
&&\left\{ A_a(\beta\varepsilon_{ap}),A_a(\beta\varepsilon_{ap})\right\}  = \lambda _T\sum_a \langle A_a(\beta\varepsilon_{ap})^2\rangle_a\,,\nonumber\\
\label{positiveness}
&&\left\{p_n B_a(\beta\varepsilon_{ap}),p_n B_a(\beta\varepsilon_{ap})\right\}  = \lambda _u\sum_a \langle p^2 B_a(\beta\varepsilon_{ap})^2\rangle_a\,,	
\end{eqnarray}
following from integral equations \eqref{AB_equations} and the definition of the total integral brackets $\{g_p,h_p\}$ \eqref{B_FullIntergalBracket} in the appendix. These brackets have the important property, $\{g_p,g_p\}\geqslant 0$ which completes the proof. It is well known from the kinetic theory that the brackets $\{g_p,h_p\}$ reflect the general properties of kinetic equations, which lead to entropy growth and the principle of kinetic coefficients symmetry (see, for example, \cite{Landau_81, FerzKap}). Thus, the developed theory shows the presence of temperature and velocity relaxation in an arbitrary two-component system for the case of small deviations of the component temperatures and velocities from their equilibrium values.

\section{Approximate solutions of the main equations of the theory \label{AppSol}}

In this section, equations \eqref{AB_equations}, \eqref{AddCond} are investigated in a $\sigma$ perturbation theory and the obtained equations in each order in $\sigma$ are solved with a truncated Sonine polynomial expansion method. The investigation is based on the following estimates of the momentum $p_{n}$ \ in $w_{ap}$
\begin{equation} \label{estimates}
{p_{n}} \sim \sqrt {{m_a}T}; \qquad m_\textrm{e}\sim \sigma^0, \qquad m_\textrm{i} \sim \sigma^{-2}.
\end{equation}
We seek the relaxation rates and the distribution functions in a $\sigma$ perturbation theory:
\begin{equation} \label{AaBa}
\lambda _T = \sum_{s \geqslant 0} \lambda _T^{(s)}, \qquad \lambda _u = \sum_{s \geqslant 0}\lambda _u^{(s)};
\qquad{A_a}\left( {\beta {\varepsilon _{ap}}} \right) = \sum\limits_{s \geqslant 0} {A_a^{(s)}\left( {\beta {\varepsilon _{ap}}} \right)},\qquad {B_a}\left( {\beta {\varepsilon _{ap}}} \right) = \sum\limits_{s \geqslant 0} {B_a^{(s)}\left( {\beta {\varepsilon _{ap}}} \right)}
\end{equation}
($A^{(s)}$ is the contribution of the order $\sigma^s$ to the quantity $A$).

\subsection {The temperature relaxation}

\subsubsection{Calculation of $A_\textrm{e}^{(0)}(x)$, $\lambda _T^{(0)}$ from equations of the order $\sigma^0$}

In the zeroth order in $\sigma$, equations \eqref{AB_equations} and additional conditions \eqref{AddCond} give the equations for the quantities $A_a^{(0)}(\beta \varepsilon _{ap})$,  $\lambda _T^{(0)}$
\[
\lambda _T^{(0)}A_\textrm{e}^{(0)}\left( {\beta {\varepsilon _{\textrm{e}p}}} \right) = {\hat K}_\textrm{ee}^{(0)}A_\textrm{e}^{(0)}\left( \beta \varepsilon_{\textrm{e}p}\right),\qquad
\lambda _T^{(0)}A_\textrm{i}^{(0)}\left(\beta \varepsilon _{\textrm{i}p} \right) = 0;
\]
\begin{equation} \label{EqT_0}
\langle A_a^{(0)}\left( \beta \varepsilon_{ap} \right)\rangle_a=0, \qquad
\langle A_\textrm{e}^{(0)}\left(\beta \varepsilon _{\textrm{e}p} \right)
{\varepsilon _{\textrm{e}p}}\rangle_\textrm{e} =\frac{3}{2}nz, \qquad
\langle A_\textrm{i}^{(0)}\left(\beta \varepsilon _{\textrm{i}p} \right)\varepsilon _{\textrm{i}p}\rangle_\textrm{i}=- \frac{3}{2}nz
\end{equation}
(see the explicit expressions for the operators $\hat K_{ab}$ in different orders in $\sigma$ in the appendix).

The first and the fourth equations \eqref{EqT_0} give the following expression for $\lambda_T^{(0)}$
\begin{equation}\label{lT_0}
\lambda_T^{(0)}=\frac{2}{3nz}\{\varepsilon_{\textrm{e}p},A_\textrm{e}^{(0)}(\beta\varepsilon_{\textrm{e}p})\}_\textrm{ee}^{(0)}=0.
\end{equation}
The zero result follows from  the definition of the integral bracket and the explicit expression for $\hat K_\textrm{ee}^{(0)} $,  because for an arbitrary function $g_p$
\begin{equation}\label{ee_0_r}
\{\varepsilon_{\textrm{e}p},g_p\}_\textrm{ee}^{(0)}=\int \rd^3p w_{\textrm{e}p} \varepsilon_{\textrm{e}p}{\hat K}_\textrm{ee}^{(0)}g_p=0.
\end{equation}

The solution of equations \eqref{EqT_0} is found in the form of a Sonine polynomial series
\begin{equation} \label{Ae0Son}
A_\textrm{e}^{(0)}(\beta \varepsilon _{\textrm{e}p}) = \sum_{m\geqslant 0} g_{\textrm{e}m}^{(0)}S_m^{1/2}(\beta\varepsilon _{\textrm{e}p} ).
\end{equation}
The chosen polynomials $S^{1/2}_m(x)$ are convenient due to their orthogonality condition
\begin{equation}
\label{Son_1/2_or}
\langle S^{1/2}_m(\beta\varepsilon_{ap})S^{1/2}_{m'}(\beta\varepsilon _{ap})\rangle_a= n_a\frac{2}{\sqrt \pi  }\frac{\Gamma (m + 3/2)}{m!}\delta_{m m'}
\end{equation}
which, in particular, gives
\begin{equation}\label{Son_1/2}
S^{1/2}_0(x)\equiv 1,\qquad S^{1/2}_1(x)=-x+3/2,\qquad S^{1/2}_2(x)=(x^2-5x+15/4)/2.
\end{equation}

{\it In matrix form}, equations \eqref{EqT_0} are given by relations
\begin{equation} \label{Eq_0}
\sum_{m'\geqslant 2} G_{\textrm{e}m,\textrm{e}m'}^{(0)}g_{\textrm{e}m'}^{(0)}=0 \qquad (m\geqslant 2);\qquad g_{\textrm{e}0}^{(0)}=0, \qquad g_{\textrm{e}1}^{(0)}=-\beta,
\end{equation}
where the matrix ${G_{am,bm'}}$ is defined by integral bracket
\begin{equation} \label{Gab}
{G_{am,bm'}} \equiv {\{ {S_m^{1/2}(\beta \varepsilon _{ap} ),S_{m'}^{1/2}( {\beta {\varepsilon _{bp}}} )} \}_{ab}}
\end{equation}
(a straightforward calculation gives $G_{\textrm{e}0,\textrm{e}m}^{(0)}=0,\,G_{\textrm{e}1,\textrm{e}m}^{(0)}=0,\, G_{\textrm{e}m,\textrm{e}0}^{(0)}=0,\, G_{\textrm{e}m,\textrm{e}1}^{(0)}=0 $ for {$\forall m$}). Equation \eqref{Eq_0} shows that $g_{\textrm{e}m}^{(0)}=0$ for $m\geqslant 2$ and, therefore,
\begin{equation} \label{Ae0}
A_\textrm{e}^{(0)}(\beta \varepsilon _{\textrm{e}p})=\beta (\beta \varepsilon _{\textrm{e}p}-3/2).
\end{equation}
It is easy to see that {\it the obtained expression \eqref{Ae0} is an exact solution of the eigenvalue problem  \eqref{EqT_0}}. The given consideration, based on the Sonine polynomial expansion, can be considered as a proof of the uniqueness of the solution. The obtained result for $A_\textrm{e}^{(0)}(\beta \varepsilon _{\textrm{i}p})$ coincides with the Landau approximation \eqref{AB_Land} \cite{Land}. However, the theory developed here allows one to find corrections of higher orders in $\sigma$ to his result.

\subsubsection{Calculation of $A_\textrm{e}^{(1)}(x)$, $A_\textrm{i}^{(0)}(x)$, $\lambda _T^{(1)}$ from equations of the order $\sigma^1$}

In the first order in $\sigma$, equations \eqref{AB_equations} and additional conditions \eqref{AddCond} give the equations for the quantities $A_\textrm{e}^{(1)}(\beta\varepsilon_{\textrm{e}p})$, $A_\textrm{i}^{(0)}(\beta\varepsilon_{\textrm{i}p})$, $\lambda_T^{(1)} $
\[
\lambda _T^{(1)}A_\textrm{e}^{(0)}\left( {\beta {\varepsilon _{\textrm{e}p}}} \right) = {\hat K}_\textrm{ee}^{(0)}A_\textrm{e}^{(1)}\left( \beta \varepsilon_{\textrm{e}p}\right),\qquad
\lambda _T^{(1)}A_\textrm{i}^{(0)}\left(\beta \varepsilon _{\textrm{i}p} \right) ={\hat K}_\textrm{ii}^{(1)}A_\textrm{i}^{(0)}\left( \beta \varepsilon_{\textrm{i}p}\right);
\]
\begin{equation} \label{EqT_1}
\langle A_\textrm{i}^{(0)}\left( \beta \varepsilon_{\textrm{i}p} \right)\rangle_\textrm{i}=0, \quad
\langle A_\textrm{i}^{(0)}\left(\beta \varepsilon _{\textrm{i}p} \right)\varepsilon _{\textrm{i}p}\rangle_\textrm{i}=- \frac{3}{2}nz; \qquad
\langle A_\textrm{e}^{(1)}\left( \beta \varepsilon_{\textrm{e}p} \right)\rangle_\textrm{e}=0, \quad
\langle A_\textrm{e}^{(1)}\left(\beta \varepsilon _{\textrm{e}p} \right)
{\varepsilon _{\textrm{e}p}}\rangle_\textrm{e} =0.
\end{equation}
Equations \eqref{EqT_1} give zero for $\lambda_T^{(1)}$
\begin{equation}\label{lT_1}
\lambda_T^{(1)}=\frac{2}{3nz}\{\varepsilon_{\textrm{e}p},A_\textrm{e}^{(1)}
(\beta\varepsilon_{\textrm{e}p})\}_\textrm{ee}^{(0)}=0,
\end{equation}
where the identity \eqref{ee_0_r} was taken into account. This expression leads to the equation ${\hat K}_\textrm{ee}^{(0)}A_\textrm{e}^{(1)}\left( \beta \varepsilon_{\textrm{e}p}\right)=0$ considered above in the main approximation but with zero additional conditions, which gives
\begin{equation} \label{Ae1}
A_\textrm{e}^{(1)}(\beta\varepsilon _{\textrm{e}p})=0.
\end{equation}
Note that the operator ${\hat K}_\textrm{ii}^{(1)}$ describes the collisions in a closed ion system [see expression \eqref{C_Kii_1} in the appendix].
The second formula in \eqref{EqT_1} with \eqref{Ae1} lead to the equation ${\hat K}_\textrm{ii}^{(1)}A_\textrm{i}^{(0)}\left( \beta \varepsilon_{\textrm{i}p}\right)=0$ and, therefore, $A_\textrm{i}^{(0)}\left( \beta \varepsilon_{\textrm{i}p} \right)$ is a hydrodynamic scalar eigenfunction of the closed ion subsystem and has the structure $c_1+c_2\varepsilon_{\textrm{i}p}$. Using additional conditions from \eqref{EqT_1} gives the constants $c_1$, $c_2$ and the result
\begin{equation} \label{Ai0}
A_\textrm{i}^{(0)}(\beta \varepsilon _{\textrm{i}p})=z\beta (3/2 - \beta \varepsilon _{\textrm{i}p}).
\end{equation}
So, {\it the spectral problem \eqref{EqT_1} is an exactly solvable one}. The result for $A_\textrm{i}^{(0)}(\beta \varepsilon _{\textrm{i}p})$ coincides with the Landau approximation \eqref{AB_Land} \cite{Land}. However, the theory developed here allows one to find corrections of higher orders in $\sigma$ to his result.

\subsubsection{Calculation of $A_\textrm{i}^{(1)}(x)$, $A_\textrm{e}^{(2)}(x)$, $\lambda _T^{(2)}$ from equations of the order $\sigma^2$}

In the second order in $\sigma$, equations \eqref{AB_equations} and additional conditions \eqref{AddCond} give the equations for the quantities $A_\textrm{e}^{(2)}(\beta\varepsilon_{\textrm{e}p})$, $A_\textrm{i}^{(1)}(\beta\varepsilon_{\textrm{i}p})$, $\lambda_T^{(2)} $
\[
\lambda _T^{(2)}A_\textrm{e}^{(0)}\left( {\beta {\varepsilon _{\textrm{e}p}}} \right) = {\hat K}_\textrm{ee}^{(0)}A_\textrm{e}^{(2)}\left( \beta \varepsilon_{\textrm{e}p}\right)+{\hat K}_\textrm{ee}^{(2)}A_\textrm{e}^{(0)}\left( \beta \varepsilon_{\textrm{e}p}\right)+{\hat K}_\textrm{ei}^{(2)}A_\textrm{i}^{(0)}\left( \beta \varepsilon_{\textrm{i}p}\right),
\]
\[
\lambda _T^{(2)}A_\textrm{i}^{(0)}\left(\beta \varepsilon _{\textrm{i}p} \right) ={\hat K}_\textrm{ie}^{(2)}A_\textrm{e}^{(0)}\left( \beta \varepsilon_{\textrm{e}p}\right)+{\hat K}_\textrm{ii}^{(1)}A_\textrm{i}^{(1)}\left( \beta \varepsilon_{\textrm{i}p}\right)+{\hat K}_\textrm{ii}^{(2)}A_\textrm{i}^{(0)}\left( \beta \varepsilon_{\textrm{i}p}\right);
\]
\begin{equation} \label{Ae2Ai1lT2}
\langle A_\textrm{e}^{(2)}(\beta\varepsilon_{\textrm{e}p})\rangle_\textrm{e}=0, \quad
\langle A_\textrm{e}^{(2)}(\beta \varepsilon _{\textrm{e}p})\varepsilon_{\textrm{e}p}\rangle_\textrm{e}=0;\quad
\langle A_\textrm{i}^{(1)}( \beta \varepsilon_{\textrm{i}p})\rangle_\textrm{i}=0, \quad
\langle A_1^{(1)}(\beta \varepsilon _{\textrm{i}p})\varepsilon _{\textrm{i}p}\rangle_\textrm{i} =0.
\end{equation}
The contributions to operators ${\hat K}_{ab}$ entering these equations are given in the appendix.

The first equation in \eqref{Ae2Ai1lT2} and additional condition for $A_\textrm{e}^{(0)}(\beta\varepsilon_{\textrm{e}p})$ from \eqref{EqT_0} give the following formula for $\lambda_T^{(2)}$
\begin{equation} \label{lT-2}
\lambda_T^{(2)}=\frac{2}{3nz}
\left(\{\varepsilon_{\textrm{e}p},A_\textrm{e}^{(0)}
(\beta\varepsilon_{\textrm{e}p})\}_\textrm{ee}^{(2)}
+\{\varepsilon_{\textrm{e}p},A_\textrm{i}^{(0)}
(\beta\varepsilon_{\textrm{i}p})\}_\textrm{ei}^{(2)}\right),
\end{equation}
where the identity \eqref{ee_0_r} was taken into account. Now, using explicit expressions for the  operators ${\hat K}_{ab}$, the definition of the integral brackets \eqref{A_IntegralBracket}, given in the appendix, and the functions $A_a^{(0)}(\beta\varepsilon_{ap})$ from \eqref{Ae0}, \eqref{Ai0}, we obtain from \eqref{lT-2}
\begin{equation} \label{lT2}
\lambda _T^{(2)} =2z^2(z+1)\sigma^2 \Lambda, \qquad \Lambda \equiv \frac{2^{5/2}\pi^{1/2}}{3}\frac{n e^4 L}{m_\textrm{e}^{1/2} T^{3/2}}\,.
\end{equation}
This expression for $\lambda _T^{(2)}$ coincides with the Landau result \cite{Land} because it is based on the functions \linebreak  $A_a^{(0)}(\beta\varepsilon_{ap})$ of his approximation \eqref{AB_Land}.

The solution of equations \eqref{Ae2Ai1lT2} is found in the form of Sonine polynomial series
\begin{equation} \label{Ae2Ai1_Son}
A_\textrm{e}^{(2)}(\beta \varepsilon _{\textrm{e}p}) = \sum_{m\geqslant 0} g_{\textrm{e}m}^{(2)}S_m^{1/2}(\beta\varepsilon _{\textrm{e}p} ), \qquad
A_\textrm{i}^{(1)}(\beta \varepsilon _{\textrm{i}p}) = \sum_{m\geqslant 0} g_{\textrm{i}m}^{(1)}S_m^{1/2}(\beta\varepsilon _{\textrm{i}p}).
\end{equation}
In a matrix form, the equations \eqref{Ae2Ai1lT2} with account for \eqref{Ae0}, \eqref{Ai0} are given by the relations
\[
\sum_{m'\geqslant 2} G_{\textrm{e}m,\textrm{e}m'}^{(0)}g_{\textrm{e}m'}^{(2)}  + z\beta G_{\textrm{e}m,\textrm{i}1}^{(2)} - \beta G_{\textrm{e}m,\textrm{e}1}^{(2)}= -\frac{3}{2} nz\beta \lambda _T^{(2)}\delta _{m1};\qquad g_{\textrm{e}0}^{(2)}=0, \quad g_{\textrm{e}1}^{(2)}=0;
\]
\begin{equation} \label{Eq_2}
\sum_{m'\geqslant 2}  G_{\textrm{i}m,\textrm{i}m'}^{(1)}{g_{\textrm{i}m'}^{(1)}}+z\beta G_{\textrm{i}m,\textrm{i}1}^{(2)} - \beta G_{\textrm{i}m,\textrm{e}1}^{(2)}=\frac{3}{2} z n\beta \lambda _T^{(2)}\delta _{m1};\qquad g_{\textrm{i}0}^{(1)}=0, \quad g_{\textrm{i}1}^{(1)}=0\,,
\end{equation}
where the matrix ${G_{am,bm'}}$ is defined in \eqref{Gab}.
The first formula from \eqref{Eq_2} gives the relations
\begin{equation}\label{explT2}
z\beta G_{\textrm{e}1,\textrm{i}1}^{(2)} - \beta G_{\textrm{e}1,\textrm{e}1}^{(2)}= -\frac{3}{2} z n\beta \lambda _T^{(2)},\qquad
\sum_{m'\geqslant 2} G_{\textrm{e}m,\textrm{e}m'}^{(0)}g_{\textrm{e}m'}^{(2)}  + z\beta G_{\textrm{e}m,\textrm{i}1}^{(2)} - \beta G_{\textrm{e}m,\textrm{e}1}^{(2)}=0 \qquad (m\geqslant 2)
\end{equation}
because substitution of \eqref{Son_1/2} into \eqref{Gab} with account for \eqref{ee_0_r} gives
\begin{equation}\label{Gee_0}
G_{\textrm{e}1,\textrm{e}m}^{(0)}=0.
\end{equation}
The first relation coincides with the expression for $\lambda_T^{(2)}$ from \eqref{lT-2}. The second one is a set of equations for the coefficients $g_{\textrm{e}m}^{(2)}\, (m\geqslant 2)$. In the one-polynomial approximation, it gives the following expression for the function $A_\textrm{e}^{(2)}\left( {\beta {\varepsilon _{\textrm{e}p}}} \right)$
\begin{equation} \label{Ae2}
A_\textrm{e}^{(2)}\left( {\beta {\varepsilon _{\textrm{e}p}}} \right) = 3\sqrt 2 z(z + 1)\beta{\sigma ^2}S_2^{1/2}( {\beta {\varepsilon _{\textrm{e}p}}} ).
\end{equation}

The fourth formula in \eqref{Eq_2} gives the relations
\begin{equation} \label{Eq_2_2}
z\beta G_{\textrm{i}1,\textrm{i}1}^{(2)} - \beta G_{\textrm{i}1,\textrm{e}1}^{(2)}=\frac{3}{2}z n\beta \lambda _T^{(2)},\qquad
\sum_{m'\geqslant 2}G_{\textrm{i}m,\textrm{i}m'}^{(1)}{g_{\textrm{i}m'}^{(1)}}=0\quad (m\geqslant 2),
\end{equation}
where the identities
\[
G_{\textrm{i}m,\textrm{i}1}^{(1)}=0\quad (m\geqslant 0); \qquad G_{\textrm{i}m,\textrm{i}1}^{(2)}=0,\qquad G_{\textrm{i}m,\textrm{e}1}^{(2)}=0\quad (m\geqslant 2)
\]
are used. They come from the explicit expressions for the components of $\hat K_{ab}$, orthogonality condition \eqref{Son_1/2_or} of the polynomials $S^{1/2}_m(x)$ and formulas \eqref{Son_1/2}. The first relation in \eqref{Eq_2_2} is an expression for $\lambda_T^{(2)}$  which is equivalent to the one from \eqref{explT2}. The second relation shows that $g_{\textrm{i}m'}^{(1)}=0\,\, (m\geqslant 2)$ and, therefore,
\begin{equation} \label{Ai1}
A_\textrm{i}^{(1)}\left( {\beta {\varepsilon _{\textrm{i}p}}} \right) = 0.
\end{equation}

\subsubsection{Calculation of $A_\textrm{e}^{(3)}(x)$, $A_\textrm{i}^{(2)}(x)$, $\lambda _T^{(3)}$ from equations of the order $\sigma^3$}

Let us start with the formula \eqref{RelaxationRates} for $\lambda_T$ using the expressions for $\hat K_{ab}$ in the appendix, which in the third order in $\sigma$ gives
\begin{equation} \label{lT_3}
\lambda_T^{(3)}=\frac{2}{3nz}\left(\{\varepsilon_{\textrm{e}p},A_\textrm{e}^{(1)}(\varepsilon_{\textrm{e}p})\}_\textrm{ee}^{(2)}
+\{\varepsilon_{\textrm{e}p},A_\textrm{e}^{(3)}(\varepsilon_{\textrm{e}p})\}_\textrm{ee}^{(0)}
+\{\varepsilon_{\textrm{e}p},A_\textrm{i}^{(1)}(\varepsilon_{\textrm{i}p})\}_\textrm{ei}^{(2)}\right)=0
\end{equation}
[see \eqref{ee_0_r}, \eqref{Ae1}, \eqref{Ai1}].

In the third order in $\sigma$, equations \eqref{AB_equations} and additional conditions \eqref{AddCond} lead to equations
\[
0= {\hat K}_\textrm{ee}^{(0)}A_\textrm{e}^{(3)}\left( \beta \varepsilon_{\textrm{e}p}\right), \qquad 0={\hat K}_\textrm{ii}^{(1)}A_\textrm{i}^{(2)}\left( \beta \varepsilon_{\textrm{i}p}\right);
\]
\begin{equation} \label{TemperatureIntegral3}
\langle A_\textrm{e}^{(3)}\left( \beta \varepsilon_{\textrm{e}p} \right)\rangle_\textrm{e}=0, \qquad
\langle A_\textrm{e}^{(3)}\left(\beta \varepsilon _{\textrm{e}p} \right)
{\varepsilon _{\textrm{e}p}}\rangle_\textrm{e} =0;\qquad
\langle A_\textrm{i}^{(2)}\left( \beta \varepsilon_{\textrm{i}p} \right)\rangle_\textrm{i}=0, \qquad
\langle A_\textrm{i}^{(2)}\left(\beta \varepsilon _{\textrm{i}p} \right)
{\varepsilon _{\textrm{i}p}}\rangle_\textrm{i} =0.
\end{equation}
The solution of the integral equation for $A_\textrm{e}^{(3)}(\beta\varepsilon_{\textrm{e}p})$ is similar to the one for $A_\textrm{e}^{(0)}(\beta\varepsilon_{\textrm{e}p})$. The solution of the integral equation for $A_\textrm{i}^{(2)}(\beta \varepsilon_{\textrm{e}p})$ is a linear combination of hydrodynamic modes of the ion subsystem. In this situation, the additional conditions give
\begin{equation} \label{Ae3Ai2}
A_\textrm{e}^{(3)}(\beta\varepsilon_{\textrm{e}p})=0, \qquad A_\textrm{i}^{(2)}(\beta\varepsilon_{\textrm{i}p})=0.
\end{equation}

\subsubsection{Calculation of $A_\textrm{i}^{(3)}(x)$, $\lambda _T^{(4)}$ from equations of the order $\sigma^4$}

Let us start with the formula \eqref{RelaxationRates} for $\lambda_T$ using expressions for $\hat K_{ab}$ in the appendix, which in the fourth order in $\sigma$ gives
\begin{equation} \label{lT_4_gen}
\lambda_T^{(4)}=\frac{2}{3nz}\left(\{\varepsilon_{\textrm{e}p},A_\textrm{e}^{(0)}(\varepsilon_{\textrm{e}p})\}_\textrm{ee}^{(4)}
+\{\varepsilon_{\textrm{e}p},A_\textrm{e}^{(2)}(\varepsilon_{\textrm{e}p})\}_\textrm{ee}^{(2)}
+\{\varepsilon_{\textrm{e}p},A_\textrm{i}^{(0)}(\varepsilon_{\textrm{i}p})\}_\textrm{ei}^{(4)}\right)
\end{equation}
[see \eqref{ee_0_r}, \eqref{Ae1}, \eqref{Ae3Ai2}]. The first and third terms here in $\lambda_T^{(4)}$ correspond to the Landau approximation \cite{Land} because they are based on the functions $A_a^{(0)}(\beta\varepsilon_{ap})$ of the Landau approximation. The calculation gives the following expression for $\lambda_T^{(4)}$
\begin{equation} \label{lT_4}
\lambda _T^{(4)} = -3z^2(z+1)\sigma^4\Lambda - 9\sqrt 2z^3(z+1)\sigma^4\Lambda.
\end{equation}
The first summand coincides with the Spitzer result based on the Landau approximation (see, for example, \cite{Ish}). The second summand takes place due to our correction $A_\textrm{e}^{(2)}(\beta\varepsilon _{\textrm{e}p})$  of the order $\sigma^2$ to the Landau contribution $A_\textrm{e}^{(0)}(\beta\varepsilon _{\textrm{e}p})$ \eqref{Ae2}.

In the fourth order in $\sigma$, equations \eqref{AB_equations} and additional conditions \eqref{AddCond} give the equations for $A_\textrm{i}^{(3)}(\beta \varepsilon_{\textrm{i}p})$
\begin{equation}
\lambda _T^{(4)}A_\textrm{i}^{(0)}(\beta \varepsilon _{\textrm{i}p})={\hat K}_\textrm{ie}^{(2)}A_\textrm{e}^{(2)}( \beta \varepsilon_{\textrm{e}p})+{\hat K}_\textrm{ie}^{(4)}A_\textrm{e}^{(0)}( \beta \varepsilon_{\textrm{e}p})+{\hat K}_\textrm{ii}^{(1)}A_\textrm{i}^{(3)}(\beta \varepsilon_{\textrm{i}p})+{\hat K}_\textrm{ii}^{(4)}A_\textrm{i}^{(0)}( \beta \varepsilon_{\textrm{i}p});
\end{equation}
\begin{equation} \label{Eq_Ai3}
\langle A_\textrm{i}^{(3)}\left(\beta \varepsilon _{\textrm{i}p} \right){\varepsilon _{\textrm{i}p}} \rangle_\textrm{i} =0.
\end{equation}
The solution of this equation in the one-polynomial approximation gives the following expression
\begin{equation} \label{Ai3}
A_\textrm{i}^{(3)}(\beta\varepsilon_{\textrm{i}p}) =2^{3/2}\beta(1+z^{-1})S_2^{1/2}(\beta\varepsilon_{\textrm{i}p})\sigma^3.
\end{equation}

\subsubsection{The temperature relaxation: results}

Finally, the above-described procedure of solution of equations in \eqref{AB_equations}, \eqref{AddCond} which are devoted to the temperature relaxation leads to the formulas
\[
A_\textrm{e}(\beta \varepsilon _{\textrm{e}p})=\beta (\beta \varepsilon _{\textrm{e}p}-3/2)+3\sqrt 2 z(z + 1)\beta{\sigma ^2}S_2^{1/2}( {\beta {\varepsilon _{\textrm{e}p}}} )+O(\sigma^3),
\]
\[
A_\textrm{i}(\beta \varepsilon _{\textrm{i}p})=z\beta (3/2 - \beta \varepsilon _{\textrm{i}p})+2^{3/2}\beta(1+z^{-1})S_2^{1/2}(\beta\varepsilon_{\textrm{i}p})\sigma^3+O(\sigma^4),
\]
\begin{equation} \label{Temp}
\lambda _T =z^2(z+1)\Lambda\sigma^2\left[2 -3(1+3z\sqrt 2)\sigma^2+O(\sigma^3)\right].
\end{equation}
This procedure can also be conducted for a higher order of the developed perturbation theory in $\sigma$.

The functions ${A_\textrm{e}}\left( {\beta {\varepsilon _{\textrm{e}p}}} \right)$  and ${A_\textrm{i}}\left( {\beta {\varepsilon _{\textrm{i}p}}} \right)$ of the order $\sigma^0$ are exact solutions of the obtained equations and they coincide with the corresponding functions \eqref{AB_Land} of the Landau approximation \cite{Land}. In his approximation, the temperature relaxation rate is given by our expression $\lambda _T^{(2)}$. We calculated the corrections $A_\textrm{e}^{(2)}(\beta \varepsilon _{\textrm{e}p})$ and $A_\textrm{i}^{(3)}(\beta\varepsilon _{\textrm{i}p})$ to the Landau results and a correction to the Spitzer expression for $\lambda _T^{(4)}$.

\subsection{The velocity relaxation}

\subsubsection{Calculation of $B_\textrm{i}^{(0)}(x)$ from equations of the order $\sigma^{-1}$}

Integral equations \eqref{AB_equations} for $B_i(\beta\varepsilon_{\textrm{i}p})$ contain on the left-hand side a momentum of the order $\sigma^{-1}$. In the order $\sigma^{-1}$, these equations and additional conditions \eqref{AddCond} give
\begin{equation}\label{IntEq_Bi0}
\lambda_u^{(0)}p_s B_\textrm{i}^{(0)}(\beta\varepsilon_{\textrm{i}p})=0;\qquad \langle B_\textrm{i}^{(0)}( \beta \varepsilon_{\textrm{i}p})\varepsilon_{\textrm{i}p}\rangle_\textrm{i}=0.
\end{equation}
The physical meaning of electron-ion collision processes shows that the attenuation constant $\lambda_u^{(0)}\neq 0$. Therefore, equations \eqref{IntEq_Bi0} show that
\begin{equation}\label{Bi0}
B_\textrm{i}^{(0)}(\beta\varepsilon_{\textrm{i}p})=0.
\end{equation}

\subsubsection{Calculation of $B_\textrm{e}^{(0)}(x)$, $B_\textrm{i}^{(1)}(x)$, $\lambda_u^{(0)}$ from equations of the order $\sigma^0$}

In the zero order in $\sigma$, equations \eqref{AB_equations} and additional conditions \eqref{AddCond} with \eqref{Bi0} give
\[
\lambda _u^{(0)}p_sB_\textrm{e}^{(0)}(\beta \varepsilon _{\textrm{e}p}) = ({\hat K}_\textrm{ee}p_s)^{(0)} B_\textrm{e}^{(0)}( \beta \varepsilon_{\textrm{e}p}),\qquad \lambda _u^{(0)}p_s B_\textrm{i}^{(1)}(\beta \varepsilon _{\textrm{e}p})=0;
\]
\begin{equation} \label{IntEq_Be0Bi1}
\langle B_\textrm{e}^{(0)}( \beta \varepsilon_{\textrm{e}p})\varepsilon_{\textrm{e}p}\rangle_\textrm{e}=\frac{3}{2}nz, \qquad
\langle B_\textrm{i}^{(1)}( \beta \varepsilon_{\textrm{i}p})\varepsilon_{\textrm{i}p}\rangle_\textrm{i}=0
\end{equation}
(explicit expressions for the operators ${\hat K}_{ab}p_s$ in different orders in $\sigma$ are given in the appendix).

The second equation in \eqref{IntEq_Be0Bi1} shows that
\begin{equation} \label{Bi1}
B_\textrm{i}^{(1)}( \beta \varepsilon_{\textrm{i}p})=0.
\end{equation}
The solution of the first equation in \eqref{IntEq_Be0Bi1} is found in the form of a Sonine polynomial series
\begin{equation}\label{Be0Son}
B_\textrm{e}^{(0)}(\beta\varepsilon _{\textrm{e}p}) = \sum_{m \geqslant 0} h_{\textrm{e}m}^{(0)}S_m^{3/2}(\beta \varepsilon _{\textrm{e}p}).
\end{equation}
The chosen polynomials $S^{3/2}_m(x)$ are convenient due to their orthogonality condition
\begin{equation}\label{g_32}
\langle S^{3/2}_m(\beta\varepsilon_{ap})S^{3/2}_{m'}(\beta\varepsilon _{ap})\beta\varepsilon _{ap}\rangle_a= n_a\frac{2}{\sqrt \pi  }\frac{\Gamma (m + 5/2)}{m!}\delta_{m m'}
\end{equation}
which, in particular, gives
\begin{equation}\label{Son_3/2}
S^{3/2}_0(x)\equiv 1,\qquad S^{3/2}_1(x)=-x+5/2,\qquad S^{3/2}_2(x)=(x^2-7x+35/4)/2.
\end{equation}
Equations \eqref{IntEq_Be0Bi1} for $B_\textrm{e}^{(0)}(\beta \varepsilon _{\textrm{e}p})$ in a matrix form are given by the relations
\begin{equation}\label{Eq_Be0lu0}
\sum_{m' \geqslant 0} H_{mm'} h_{\textrm{e}m'}^{(0)} =0,\qquad h_{\textrm{e}0}^{(0)}=\beta,
\end{equation}
where the notation
\begin{equation} \label{Hmm'}
H_{mm'}=H_{\textrm{e}m,\textrm{e}m'}^{(0)}-\lambda _u^{(0)}n z m_\textrm{e} \beta^{-1} \frac{4}{\sqrt \pi  }\frac{\Gamma(m + 5/2)}{{m!}}\delta_{mm'}
\end{equation}
and the integral brackets $H_{am,bm'}$:
\begin{equation} \label{H_Brackets}
H_{am,bm'} \equiv \left\{p_s S_m^{3/2}(\beta\varepsilon_{ap}),p_s S_{m'}^{3/2}(\beta\varepsilon_{bp})\right\}_{ab}
\end{equation}
are introduced.
Equations \eqref{Eq_Be0lu0} in the one-polynomial approximation give the following result
\begin{equation}\label{Be0lu0}
B_\textrm{e}^{(0)}(\beta\varepsilon _{\textrm{e}p}) = \beta,\qquad \lambda_u^{(0)} = z^2\Lambda,
\end{equation}
where the quantity $\Lambda$ is defined in \eqref{lT-2}. These expressions coincide with the result obtained in the Landau approximation \eqref{AB_Land} (see, for example, \cite{AlBogdRukh}). The theory developed in the present paper allows one to find corrections of higher orders in $\sigma$ to this result.

\subsubsection{Calculation of $B_\textrm{e}^{(1)}(x)$, $B_\textrm{i}^{(2)}(x)$, $\lambda_u^{(1)}$ from equations of the order $\sigma^1$}

In the first order in $\sigma$, equations \eqref{AB_equations} and additional conditions \eqref{AddCond} with account for the explicit expressions for $\hat K_{ab} p_s $ in the appendix and \eqref{Bi0} give the equations for $B_\textrm{e}^{(1)}(\beta\varepsilon_{\textrm{e}p})$, $\lambda_u^{(1)}$, $B_\textrm{i}^{(2)}(\beta\varepsilon_{\textrm{i}p})$
\[
\lambda _u^{(1)}p_s B_\textrm{e}^{(0)}(\beta \varepsilon _{\textrm{e}p})+\lambda _u^{(0)}p_s B_\textrm{e}^{(1)}(\beta \varepsilon _{\textrm{e}p}) = ({\hat K}_\textrm{ee}p_s)^{(0)} B_\textrm{e}^{(1)}( \beta \varepsilon_{\textrm{e}p}),\qquad \langle B_\textrm{e}^{(1)}( \beta \varepsilon_{\textrm{e}p})\varepsilon_{\textrm{e}p}\rangle_\textrm{e}=0;
\]
\begin{equation} \label{IntEq_Be1Bi2}
\lambda _u^{(0)}p_s B_\textrm{i}^{(2)}(\beta \varepsilon _{\textrm{i}p})= ({\hat K}_\textrm{ie}p_s)^{(1)} B_\textrm{e}^{(0)}( \beta \varepsilon_{\textrm{e}p}), \qquad \langle B_\textrm{i}^{(2)}( \beta \varepsilon_{\textrm{i}p})\varepsilon_{\textrm{i}p}\rangle_\textrm{i}=-\frac{3}{2}nz\sigma^2.
\end{equation}
The solution of the first equation in \eqref{IntEq_Be1Bi2} is found in the form a Sonine polynomial series
\begin{equation}\label{Be1_Son}
B_\textrm{e}^{(1)}(\beta\varepsilon _{\textrm{e}p}) = \sum_{m \geqslant 0} {h_{\textrm{e}m}^{(1)}S_m^{{3/2}}(\beta\varepsilon _{\textrm{e}p})}.
\end{equation}
The equations for $B_\textrm{e}^{(1)}(\beta\varepsilon _{\textrm{e}p})$  in \eqref{IntEq_Be1Bi2} in a matrix notation have the form
\begin{equation}
\sum_{{m'} \geqslant 0} H_{mm'}{h_{\textrm{e}m'}^{(1)}} =\lambda_u^{(1)}3n z m_\textrm{e} \delta _{m0}\,,\qquad {h_{\textrm{e}0}^{(1)}}=0
\end{equation}
which gives
\begin{equation}\label{he_1}
\sum_{{m'} \geqslant 1} H_{\textrm{e}0,\textrm{e}m'}^{(0)}{h_{\textrm{e}m'}^{(1)}} =\lambda_u^{(1)}3n z m_\textrm{e}\,,
\qquad \sum_{{m'} \geqslant 1} H_{mm'}{h_{\textrm{e}m'}^{(1)}} =0\qquad (m\geqslant 1).
\end{equation}
The second equation here has only a trivial solution, therefore, formulas \eqref{IntEq_Be1Bi2}, \eqref{Be1_Son}  lead to the expressions
\begin{equation}
B_\textrm{e}^{(1)}\left( {\beta {\varepsilon _{\textrm{e}p}}} \right) = 0, \qquad  \lambda_u^{(1)}=0.
\end{equation}
The third and the fourth relations in \eqref{IntEq_Be1Bi2} with account for expressions \eqref{Be0lu0} obtained in the one-polynomial approximation give
\begin{equation}\label{Bi2}
B_\textrm{i}^{(2)}(\beta\varepsilon_{\textrm{i}p})=-z\beta \sigma^2.
\end{equation}
According to \eqref{AB_Land}, this expression corresponds to the Landau approximation.

\subsubsection{Calculation of $B_\textrm{e}^{(2)}(x)$, $B_\textrm{i}^{(3)}(x)$, $\lambda_u^{(2)}$ from equations of the order $\sigma^2$}

In the second order in $\sigma$, equations \eqref{AB_equations} and additional conditions \eqref{AddCond} with account for the expressions for $\hat K_{ab} p_s $  in the appendix and \eqref{Bi0}  give equations for the quantities $B_\textrm{e}^{(2)}(\beta\varepsilon_{\textrm{e}p})$, $\lambda_u^{(2)}$, $B_\textrm{i}^{(3)}(\beta\varepsilon_{\textrm{i}p})$
\[
\lambda _u^{(2)}p_s B_\textrm{e}^{(0)}(\beta \varepsilon _{\textrm{e}p})+\lambda _u^{(0)}p_s B_\textrm{e}^{(2)}(\beta \varepsilon _{\textrm{e}p}) = ({\hat K}_\textrm{ee}p_s)^{(2)} B_\textrm{e}^{(0)}( \beta \varepsilon_{\textrm{e}p})+({\hat K}_\textrm{ee}p_s)^{(0)} B_\textrm{e}^{(2)}( \beta \varepsilon_{\textrm{e}p})+({\hat K}_\textrm{ei}p_s)^{(0)} B_\textrm{i}^{(2)}( \beta \varepsilon_{\textrm{i}p}),
\]
\[
\langle B_\textrm{e}^{(2)}( \beta \varepsilon_{\textrm{e}p})\varepsilon_{\textrm{e}p}\rangle_\textrm{e}=0;
\]
\begin{equation} \label{IntEq_Be2Bi3}
\lambda _u^{(0)}p_s B_\textrm{i}^{(3)}(\beta \varepsilon _{\textrm{i}p})= ({\hat K}_\textrm{ii}p_s)^{(0)} B_\textrm{i}^{(2)}( \beta \varepsilon_{\textrm{i}p}), \qquad \langle B_\textrm{i}^{(3)}( \beta \varepsilon_{\textrm{i}p})\varepsilon_{\textrm{i}p}\rangle_\textrm{i}=0.
\end{equation}
The third relation here with the explicit expression for $({\hat K}_\textrm{ii}p_s)^{(0)}$ shows that
\begin{equation}\label{Bi3}
B_\textrm{i}^{(3)}(\beta \varepsilon _{\textrm{i}p})=0.
\end{equation}
We seek the solution of the first equation in \eqref{IntEq_Be2Bi3} as a Sonine polynomial series
\begin{equation}\label{Be2_Son}
B_\textrm{e}^{(2)}(\beta\varepsilon _{\textrm{e}p}) = \sum_{m \geqslant 0} {h_{\textrm{e}m}^{(2)}S_m^{{3/2}}(\beta\varepsilon _{\textrm{e}p})}.
\end{equation}
The first and the second equations in \eqref{IntEq_Be2Bi3} in a matrix form with the notation \eqref{Hmm'} are given by
\begin{equation} \label{PolynimialEquationForSigmaV2}
\sum_{m' \geqslant 0}  H_{mm'}{h_{\textrm{e}m'}^{(2)}}=\lambda _u^{(2)}3nz m_\textrm{e} \delta _{m0}+ z\beta\sigma ^2H_{\textrm{e}m,\textrm{i}0}^{(0)} - \beta H_{\textrm{e}m,\textrm{e}0}^{(2)} ,\qquad h_{\textrm{e}0}^{(2)}=0
\end{equation}
which leads to the relations
\[
\sum_{{m'} \geqslant 1} H_{\textrm{e}0,\textrm{e}m'}^{(0)}{h_{\textrm{e}m'}^{(2)}} =\lambda_u^{(2)}3n z m_\textrm{e}+  z\beta\sigma ^2H_{\textrm{e}0,\textrm{i}0}^{(0)} - \beta H_{\textrm{e}0,\textrm{e}0}^{(2)},
\]
\begin{equation}\label{he_2}
\sum_{{m'} \geqslant 1} H_{mm'}{h_{\textrm{e}m'}^{(2)}} = z\beta\sigma ^2H_{\textrm{e}m,\textrm{i}0}^{(0)} - \beta H_{\textrm{e}m,\textrm{e}0}^{(2)}\qquad (m\geqslant 1).
\end{equation}
The second relation here is a set of equations for the coefficients $h_{\textrm{e}m}^{(2)}$, the first equation allows one to calculate $\lambda_u^{(2)}$. In the one-polynomial approximation, they lead to the expressions for $B_\textrm{e}^{(2)}(\beta\varepsilon_{\textrm{e}p})$
\begin{equation}\label{Be2}
B_\textrm{e}^{(2)}(\beta\varepsilon_{\textrm{e}p}) = -\frac{3z(2z - 1)}{3z + 4\sqrt 2 }\beta\left(5/2 - \beta \varepsilon _{\textrm{e}p}\right)\sigma ^2,
\end{equation}
and $\lambda_u^{(2)}$
\begin{equation}\label{lu2}
\lambda_u^{(2)} =\frac{1}{2}z^2(2z-1)\sigma^2\Lambda-\frac{9}{2}\frac{z^3(2z-1)}{3z+2^{5/2}}\sigma^2\Lambda.
\end{equation}

\subsubsection{Calculation of $B_\textrm{e}^{(3)}(x)$, $B_\textrm{i}^{(4)}(x)$, $\lambda_u^{(3)}$ from equations of the order $\sigma^3$}

In the third order in $\sigma$, equations \eqref{AB_equations} and additional conditions \eqref{AddCond} with account for the explicit expressions for $\hat K_{ab} p_s $ in the appendix and \eqref{Bi0} give the equations for the quantities $B_\textrm{e}^{(3)}(\beta\varepsilon_{\textrm{e}p})$, $\lambda_u^{(3)}$, $B_\textrm{i}^{(4)}(\beta\varepsilon_{\textrm{i}p})$
\[
\lambda _u^{(3)}p_s B_\textrm{e}^{(0)}(\beta \varepsilon _{\textrm{e}p})+\lambda _u^{(0)}p_s B_\textrm{e}^{(3)}(\beta \varepsilon _{\textrm{e}p}) = ({\hat K}_\textrm{ee}p_s)^{(0)} B_\textrm{e}^{(3)}( \beta \varepsilon_{\textrm{e}p}),\qquad \langle B_\textrm{e}^{(3)}( \beta \varepsilon_{\textrm{e}p})\varepsilon_{\textrm{e}p}\rangle_\textrm{e}=0;
\]
\[
\lambda _u^{(0)}p_s B_\textrm{i}^{(4)}(\beta \varepsilon _{\textrm{i}p})+\lambda _u^{(2)}p_s B_\textrm{i}^{(2)}(\beta \varepsilon _{\textrm{i}p})= ({\hat K}_\textrm{ii}p_s)^{(1)} B_\textrm{i}^{(2)}( \beta \varepsilon_{\textrm{i}p})+
 ({\hat K}_\textrm{ie}p_s)^{(3)} B_\textrm{e}^{(0)}( \beta \varepsilon_{\textrm{e}p})+({\hat K}_\textrm{ie}p_s)^{(1)} B_\textrm{e}^{(2)}( \beta \varepsilon_{\textrm{e}p}),
\]
\begin{equation} \label{IntEq_Be3Bi4}
\langle B_\textrm{i}^{(4)}( \beta \varepsilon_{\textrm{i}p})\varepsilon_{\textrm{i}p}\rangle_\textrm{i}=0.
\end{equation}
By analogy with the calculation of $B_\textrm{e}^{(1)}(\beta\varepsilon_{\textrm{e}p})$, $\lambda_u^{(1)}$, the first equation here gives
\begin{equation} \label{Be3lu3}
B_\textrm{e}^{(3)}\left( {\beta {\varepsilon _{\textrm{e}p}}} \right) = 0, \qquad \lambda _u^{(3)}=0.
\end{equation}
The second equation in \eqref{IntEq_Be3Bi4} with account for the above calculated quantities in the one-polynomial approximation gives
\begin{equation}\label{Bi4}
B_\textrm{i}^{(4)}\left( {\beta {\varepsilon _{\textrm{i}p}}} \right) =  - \frac{{3}}{{5}}z\beta\left( {5/2 - \beta {\varepsilon _{\textrm{i}p}}} \right){\sigma ^4}.
\end{equation}

\subsubsection{The velocity relaxation: results}

Finally, the above-described procedure of the solution of the equations in \eqref{AB_equations}, \eqref{AddCond} which are devoted to the velocity relaxation leads to the formulas
\[
B_\textrm{e}(\beta\varepsilon_{\textrm{e}p}) = \beta - \frac{3z(2z - 1)}{3z + 4\sqrt 2 }\beta\left(5/2 - \beta \varepsilon _{\textrm{e}p}\right)\sigma ^2 + O(\sigma ^3),
\]
\[
B_\textrm{i}(\beta\varepsilon_{\textrm{i}p}) =  - z\beta\sigma^2 - \frac{3}{5}z\beta\left(5/2 - \beta\varepsilon _{\textrm{i}p} \right)\sigma ^4 + O(\sigma ^5),
\]
\begin{equation} \label{lu}
{\lambda _u} = {z^2}\Lambda\left[1  + \frac{1}{2}\left( {2z - 1} \right){\sigma ^2}  - \frac{9}{2}\frac{z(2z - 1)}{{3z + {2^{{5 \mathord{\left/
{\vphantom {5 2}} \right.
 \kern-\nulldelimiterspace} 2}}}}}{\sigma ^2}  + O\left( {{\sigma ^3}} \right)\right]
={z^2}\Lambda\left[1  + \left( {2z - 1} \right)\frac{{{2^{{3 \mathord{\left/
{\vphantom {3 2}} \right.
\kern-\nulldelimiterspace} 2}}} - 3z}}{{3z + {2^{{5 \mathord{\left/
{\vphantom {5 2}} \right.
\kern-\nulldelimiterspace} 2}}}}}{\sigma ^2} + O\left( {{\sigma ^3}} \right)
\right].
\end{equation}
This procedure can be conducted for a higher order of the developed perturbation theory in $\sigma$.

The functions $B_\textrm{e}^{(0)}\left( {\beta {\varepsilon _{\textrm{e}p}}} \right)$ and $B_\textrm{i}^{(2)}\left( {\beta {\varepsilon _{\textrm{i}p}}} \right)$ given by \eqref{Be0lu0}, \eqref{Bi2} were calculated in the one-polynomial approximation from integral equations \eqref{AB_equations}. They coincide with the ones that follow from the Landau approximation \eqref{AB_Land}. Our theory gives corrections to these functions of the order $\sigma^2$ and $\sigma^4$, respectively. These corrections are calculated in the one-polynomial approximation. In the principal order in $\sigma$, the expression for $\lambda_u$ \eqref{Be0lu0} coincides with the result \cite{AlBogdRukh} obtained in the Landau approximation. Our theory gives a correction to this result of the order $\sigma^2$. The first term in the contribution \eqref{lu2} to the velocity relaxation rate $\lambda_u^{(2)}$ [the second term in \eqref{lu}] can be obtained in the Landau approximation by an additional expansion in $\sigma$ of the Landau collision integral, the second term [the third term in \eqref{lu}] is due to the account for our correction $B_\textrm{e}^{(2)}\left( {\beta {\varepsilon _{\textrm{e}p}}} \right)$ of the order $\sigma^2$ to $B_\textrm{e}^{(0)}\left( {\beta {\varepsilon _{\textrm{e}p}}} \right)$.

\section{Conclusion}

The relaxation of the temperatures and velocities of the components of a quasi-equilibrium two-component homogeneous fully ionized plasma described by the Landau kinetic equation is investigated.

The Chapman-Enskog method is generalized to take into account the kinetic modes of the system. The generalization was made on the basis of the idea of Bogoliubov functional hypothesis.

In the approximation of a small difference of the component temperatures and velocities it is shown that relaxation really exists (the relaxation rates are positive). This proof is based on the arguments that are valid for an arbitrary two-component system because they rely on the general properties of kinetic equations.

The integral equations for the functions $A_a(\beta\varepsilon_{ap})$ describing the temperature kinetic mode of the system are solved approximately in a $\sigma=(m_\textrm{e}/m_\textrm{i})^{1/2}$ perturbation theory (i.e., in the small electron-to-ion mass ratio) up to the fourth order in $\sigma$. The equations of each order are solved with the help of a truncated expansion in the Sonine polynomials.

It is shown that the equations for the zero order in $\sigma$ contributions $A_a^{(0)}(\beta\varepsilon_{ap})$ are exactly solvable and $A_a^{(0)}(\beta\varepsilon_{ap})$ coincide with the distribution functions of the Landau theory \cite{Land}. The developed theory gives the main corrections $A_\textrm{e}^{(2)}(\beta\varepsilon_{\textrm{e}p})$,  $A_\textrm{i}^{(3)}(\beta\varepsilon_{\textrm{i}p})$ of the orders $\sigma^2$ and $\sigma^3$ to the functions $A_a^{(0)}(\beta\varepsilon_{ap})$. The corrections are calculated in the one-polynomial approximation. The temperature relaxation rate $\lambda_T$ is found. The main contribution $\lambda_T^{(2)}$ to $\lambda_T$ coincides with the Landau result. The developed theory corrects the $\sigma^4$ term $\lambda_T^{(4)}$ of the Spitzer result for $\lambda_T$ due to the account for our correction $A_\textrm{e}^{(2)}(\beta\varepsilon_{\textrm{e}p})$ to the function $A_\textrm{e}^{(0)}(\beta\varepsilon_{\textrm{e}p})$.  The Spitzer contribution to $\lambda_T^{(4)}$ is related to the Landau functions $A_a^{(0)}(\beta\varepsilon_{ap})$ with an additional expansion in $\sigma$-powers of the Landau collision integral.

The integral equations for the functions $B_a(\beta\varepsilon_{ap})$ describing the velocity kinetic mode of the system are solved approximately in a $\sigma$ perturbation theory. The main in $\sigma$ contributions $B_\textrm{e}^{(0)}(\beta\varepsilon_{\textrm{e}p})$,  $B_\textrm{i}^{(2)}(\beta\varepsilon_{\textrm{i}p})$ to $B_a(\beta\varepsilon_{ap})$
are calculated in the one-polynomial approximation. They coincide with the results of the theory \cite{AlBogdRukh} based on the Landau approximation. The developed theory gives the main corrections $B_\textrm{e}^{(2)}(\beta\varepsilon_{\textrm{e}p})$, $B_\textrm{i}^{(4)}(\beta\varepsilon_{\textrm{i}p})$ to $B_\textrm{e}^{(0)}(\beta\varepsilon_{\textrm{e}p})$, $B_\textrm{i}^{(2)}(\beta\varepsilon_{\textrm{i}p})$, respectively. The corrections are calculated in the one-polynomial approximation. The velocity relaxation rate $\lambda_u$ is obtained. The principal order in $\sigma$ contribution $\lambda_u^{(0)}$ calculated in the one-polynomial approximation coincides with the result of the theory \cite{AlBogdRukh} based on the Landau approximation. The developed theory gives the correction $\lambda_u^{(2)}$ of the order $\sigma^2$ to this result calculated in the one-polynomial approximation too.

The obtained results show that in the hydrodynamic theory of the system, the violation of local equilibrium in the presence of relaxation processes should be taken into account.

%\section*{Appendix}

\appendix

\section{The integral brackets and the linearized collision operator}

The integral bracket $\{g_p,h_p\}_{ab}$ for arbitrary functions $g_p$, $h_p$  is defined by the formula
\begin{equation} \label{A_IntegralBracket}
\{g_p,h_p\}_{ab}=-\int \rd^3 p d^3 p' M_{ab}(p,p')w_{bp'}g_p h_{p'}\,,
\end{equation}
where the kernel $M_{ab}(p,p')$ of this form is given by the linearized Landau collision integral \eqref{LandauCI1}. To simplify the calculations, it is convenient to introduce the linearized collision operator ${\hat K}_{ab}$
\begin{equation}\label{A_Kab_def}
{\hat K}_{ab}h_p=\int \rd^3p'K_{ab}(p,p')h_{p'}\,,\qquad
w_{ap}K_{ab}(p,p')\equiv - M_{ab}(p,p')w_{bp'}
\end{equation}
which allows one to rewrite the form \eqref{A_IntegralBracket} as
\[
\{g_p,h_p\}_{ab}=\langle g_p{\hat K}_{ab}h_p\rangle_a
\]
[see \eqref{av}]. An explicit expression for the operator ${\hat K}_{ab}$ is given by the formula
\begin{equation} \label{A_ExplicitIntegralBracket}
{\hat K}_{ab} h_p = 2\pi e_a^2 L\,w_{ap}^{-1} \frac{\partial}{\partial p_n} \sum_c e_c^2 \int \rd^3p'\left(\delta_{cb}\frac{\partial h_{p'}}{\partial p'_{l}}-\delta_{ab}\frac{\partial h_p}{\partial p_l}  \right)D_{nl}\left(\frac{p}{m_a}-\frac{p'}{m_c}\right)w_{ap}w_{cp'}
\end{equation}
following from the Landau collision integral \eqref{LandauCI}.

Bilinear form \eqref{A_IntegralBracket} has important properties
\begin{equation}\label{A_r_prop}
\sum_a\{\varepsilon_{ap},g_p\}_{ab}=0, \qquad \sum_a\{p_n,g_p\}_{ab}=0, \qquad \{1,g_p\}_{ab}=0;
\end{equation}
\begin{equation}\label{A_l_prop}
\sum_b\{g_p,\varepsilon_{bp}\}_{ab}=0, \qquad \sum_b\{g_p,p_n\}_{ab}=0, \qquad \{g_p,1\}_{ab}=0.
\end{equation}
Formulas \eqref{A_r_prop} follow from the relations \eqref{TotalSources}, \eqref{CI_property}; formulas \eqref{A_l_prop} are true because the Maxwell distribution is an equilibrium distribution: $\left.I_{ap}(f)\right|_{f_{bp}= w_{b,p-m_b\upsilon}(T)}=0$.

An important tool of our investigation is the total integral bracket defined by the formula
\begin{equation} \label{B_FullIntergalBracket}
\{g_p,h_p\} \equiv \sum_{a,b} \{g_p,h_p\}_{ab}	\,,	
\end{equation}
where the forms $\{g_p,h_p\}_{ab}$ are given by relation \eqref{A_IntegralBracket}.
A simple calculation leads to the relation
\begin{equation} \label{A_ExplicitFullIntergalBracket}
\{g_p,h_p\} = \pi L\sum\limits_{a,b} e_a^2e_b^2 \int \rd^3 p \, d^3 p'  D_{nl}\left(\frac{p}{m_a}-\frac{p'}{m_b}\right) w_{ap}w_{bp'}\left( \frac{\partial g_p}{\partial p_n} - \frac{\partial g_{p'}}{\partial p'_n}\right)\left(\frac{\partial h_p}{\partial p_l} - \frac{\partial h_{p'}}{\partial p'_l} \right)	
\end{equation}
which shows that this form has the properties
\begin{equation} \label{B_FullBracketProperties}
\{g_p,g_p\} \geqslant 0, \qquad \{g_p,h_p\}=\{h_p,g_p\}.
\end{equation}

\section{Explicit expressions for the linearized collision operator in the \\ $\sigma$-perturbation theory}

The explicit expressions for the functions ${\hat K_{ab}}{A_b}\left( {\beta {\varepsilon _{bp}}} \right)$ are
\[
\sum_b {\hat K}_{\textrm{e}b} {A_b}\left( {\beta {\varepsilon _{bp}}} \right) = 2\pi {e^4}m_\textrm{e} L w_{\textrm{e}p}^{-1}\frac{\partial }{{\partial {p_n}}} \int {\rd^3}{p'} {{w_{\textrm{e}p}}{w_{\textrm{e}p'}}{D_{nl}(p-p')}\left[ {\frac{{\partial {A_\textrm{e}}\left( {\beta {\varepsilon _{ep'}}} \right)}}{{\partial {p'_l}}} - \frac{{\partial {A_\textrm{e}}\left( {\beta {\varepsilon _{\textrm{e}p}}} \right)}}{{\partial {p_l}}}} \right]}
\]
\[
+ 2\pi {z^2}{e^4}m_\textrm{e} L w_{\textrm{e}p}^{-1}\frac{\partial }{{\partial {p_n}}}\int {\rd^3}{p'} {{w_{\textrm{e}p}}{w_{\textrm{i}p'}}{D_{nl}(p-p'\sigma^2)}\left[ {\frac{{\partial {A_\textrm{i}}\left( {\beta {\varepsilon _{\textrm{i}p'}}} \right)}}{{\partial {p'_l}}} - \frac{{\partial {A_\textrm{e}}\left( {\beta {\varepsilon _{\textrm{e}p}}} \right)}}{{\partial {p_l}}}} \right]},
\]
\[
\sum_b {\hat K}_{\textrm{i}b} {A_b}\left( {\beta {\varepsilon _{bp}}} \right) = 2\pi {z^2}{e^4}m_\textrm{e} L w_{\textrm{i}p}^{-1}\frac{\partial }{{\partial {p_n}}}\int {\rd^3}{p'} {{w_{\textrm{i}p}}{w_{\textrm{e}p'}}{D_{nl}(p'-p\sigma^2)}\left[ {\frac{{\partial {A_\textrm{e}}\left( {\beta {\varepsilon _{\textrm{e}p'}}} \right)}}{{\partial {p'_l}}} - \frac{{\partial {A_\textrm{i}}\left( {\beta {\varepsilon _{\textrm{i}p}}} \right)}}{{\partial {p_l}}}} \right]}
\]
\begin{equation} \label{C_KabAb}
+ 2\pi {z^4}{e^4}m_\textrm{i} L w_{\textrm{i}p}^{-1}\frac{\partial }{{\partial {p_n}}}\int {\rd^3}{p'} {{w_{\textrm{i}p}}{w_{\textrm{i}p'}}{D_{nl}(p-p')}\left[ {\frac{{\partial {A_\textrm{i}}\left( {\beta {\varepsilon _{\textrm{i}p'}}} \right)}}{{\partial {p'_l}}} - \frac{{\partial {A_\textrm{i}}\left( {\beta {\varepsilon _{\textrm{i}p}}} \right)}}{{\partial {p_l}}}} \right]}.
\end{equation}

These expressions show that expansion in the $\sigma$-powers of the operators ${\hat K}_{ab}$ acting  \emph{on the even functions} has the structure
\[
{\hat K}_\textrm{ee}^{}={\hat K}_\textrm{ee}^{(0)}+{\hat K}_\textrm{ee}^{(2)}+O(\sigma^4), \qquad
{\hat K}_\textrm{ei}^{}={\hat K}_\textrm{ei}^{(2)}+{\hat K}_\textrm{ei}^{(4)}+O(\sigma^6),
\]
\begin{equation}\label{C_K_even}
{\hat K}_\textrm{ie}^{}={\hat K}_\textrm{ie}^{(2)}+{\hat K}_\textrm{ie}^{(4)}+O(\sigma^6), \qquad
{\hat K}_\textrm{ii}^{}={\hat K}_\textrm{ii}^{(1)}+{\hat K}_\textrm{ii}^{(2)}+O(\sigma^4).
\end{equation}
The explicit expressions for the operators entering here are
\begin{align}
%\[
{\hat K}_\textrm{ee}^{(0)} g_p &= 2\pi {e^4}m_\textrm{e} L w_{\textrm{e}p}^{-1}\frac{\partial }{{\partial {p_n}}} \int {\rd^3}{p'} {{w_{\textrm{e}p}}{w_{\textrm{e}p'}}{D_{nl}(p-p')}\left( {\frac{{\partial g_{p'}}}{{\partial {p'_l}}} - \frac{{\partial g_p}}{{\partial {p_l}}}} \right)}  \nonumber\\
%\]
%\begin{equation}
\label{C_Kee0}
&\phantom{=}- 2\pi {z^2}{e^4}m_\textrm{e} L n w_{\textrm{e}p}^{-1}\frac{\partial }{{\partial {p_n}}}{w_{\textrm{e}p}}
D_{nl}(p) \frac{\partial g_p}{\partial {p_l}},\\
%\end{equation}
%\begin{equation}
\label{C_Kii_1}
{\hat K}_\textrm{ii}^{(1)}g_p &= 2\pi {z^4}{e^4}m_\textrm{i} L w_{\textrm{i}p}^{-1}\frac{\partial }{{\partial {p_n}}}\int {\rd^3}{p'} {{w_{\textrm{i}p}}{w_{\textrm{i}p'}}{D_{nl}(p-p')}\left( {\frac{{\partial g_{p'}}}{{\partial {p'_l}}} - \frac{{\partial g_p}}{{\partial {p_l}}}} \right)},\\
%\end{equation}
%\begin{equation}
\label{C_Kie_2}
{\hat K}_\textrm{ie}^{(2)}g_p &=- 2\pi {z^2}{e^4}m_\textrm{e}\sigma^2 L w_{\textrm{i}p}^{-1}\frac{\partial }{\partial {p_n}}\int \rd^3 p' w_{\textrm{i}p}w_{\textrm{e}p'}p_s\frac{\partial D_{nl}(p')}{\partial p'_s} \frac{\partial g_{p'}}{\partial {p'_l}},\\
%\end{equation}
%\begin{equation}
\label{C_Kei_2}
{\hat K}_\textrm{ei}^{(2)}g_p & = -2\pi {z^2}{e^4}m_\textrm{e}\sigma^2 L w_{\textrm{e}p}^{-1}\frac{\partial }{\partial {p_n}}\int \rd^3 p' w_{\textrm{e}p}w_{\textrm{i}p'}p'_s\frac{\partial D_{nl}(p)}{\partial p_s} \frac{\partial g_{p'}}{\partial {p'_l}},\\
%\end{equation}
%\[
{\hat K}_\textrm{ee}^{(2)}g_p&=- 2\pi {z^2}{e^4}m_\textrm{e}\sigma^4 L w_{\textrm{e}p}^{-1}\frac{\partial }{\partial {p_n}}\int \rd^3 p' w_{\textrm{e}p}w_{\textrm{i}p'}p'_s p'_r\frac{\partial^2 D_{nl}(p)}{\partial p_s \partial p_r} \frac{\partial g_{p}}{\partial {p_l}} \nonumber\\
%\]
%\begin{equation}
\label{C_Kee_2}
&=- 2\pi {z^2}{e^4}nL\,T \sigma^2m_\textrm{e}^2 w_{\textrm{e}p}^{-1}\frac{\partial }{\partial {p_n}} w_{\textrm{e}p}\frac{\partial^2 D_{nl}(p)}{\partial p_s \partial p_s} \frac{\partial g_{p}}{\partial {p_l}},\\
%\end{equation}
%\begin{equation}
\label{C_Kii_2}
{\hat K}_\textrm{ii}^{(2)}g_p &= -2\pi {z^2}{e^4}m_\textrm{e} L w_{\textrm{i}p}^{-1}\frac{\partial }{\partial {p_n}}\int \rd^3 p' w_{\textrm{i}p}w_{\textrm{e}p'}D_{nl}(p') \frac{\partial g_{p}}{\partial {p_l}}=- z^3 T m_\textrm{e} \Lambda w_{\textrm{i}p}^{-1}\frac{\partial }{\partial {p_n}} w_{\textrm{i}p} \frac{\partial g_{p}}{\partial {p_n}}.
%\end{equation}
\end{align}

The explicit expressions for the functions ${\hat K_{ab}}{p_s}{B_b}\left( {\beta {\varepsilon _{bp}}} \right)$ are
\[
\sum_b {\hat K}_{\textrm{e}b} p_s{B_b}\left( {\beta {\varepsilon _{bp}}} \right) = 2\pi {e^4}m_\textrm{e} L w_{\textrm{e}p}^{-1}\frac{\partial }{{\partial {p_n}}} \int {\rd^3}{p'} {{w_{\textrm{e}p}}{w_{\textrm{e}p'}}{D_{nl}(p-p')}\left[ {\frac{{\partial p'_s{B_\textrm{e}}\left( {\beta {\varepsilon _{\textrm{e}p'}}} \right)}}{{\partial {p'_l}}} - \frac{{\partial p_s{B_\textrm{e}}\left( {\beta {\varepsilon _{\textrm{e}p}}} \right)}}{{\partial {p_l}}}} \right]}
\]
\[
+ 2\pi {z^2}{e^4}m_\textrm{e} L w_{\textrm{e}p}^{-1}\frac{\partial }{{\partial {p_n}}}\int {\rd^3}{p'} {{w_{\textrm{e}p}}{w_{\textrm{i}p'}}{D_{nl}(p-p'\sigma^2)}\left[ {\frac{{\partial p'_s{B_\textrm{i}}\left( {\beta {\varepsilon _{\textrm{i}p'}}} \right)}}{{\partial {p'_l}}} - \frac{{\partial p_s{B_\textrm{e}}\left( {\beta {\varepsilon _{\textrm{e}p}}} \right)}}{{\partial {p_l}}}} \right]},
\]
\[
\sum_b {\hat K}_{\textrm{i}b} p_s{B_b}\left( {\beta {\varepsilon _{bp}}} \right) = 2\pi {z^2}{e^4}m_\textrm{e} L w_{\textrm{i}p}^{-1}\frac{\partial }{{\partial {p_n}}}\int {\rd^3}{p'} {{w_{\textrm{i}p}}{w_{\textrm{e}p'}}{D_{nl}(p'-p\sigma^2)}\left[ {\frac{{\partial p'_s{B_\textrm{e}}\left( {\beta {\varepsilon _{\textrm{e}p'}}} \right)}}{{\partial {p'_l}}} - \frac{{\partial p_s{B_\textrm{i}}\left( {\beta {\varepsilon _{\textrm{i}p}}} \right)}}{{\partial {p_l}}}} \right]}  \]
\begin{equation} \label{C_KabBb}
+ 2\pi {z^4}{e^4}m_\textrm{i} L w_{\textrm{i}p}^{-1}\frac{\partial }{{\partial {p_n}}}\int {\rd^3}{p'} {{w_{\textrm{i}p}}{w_{\textrm{i}p'}}{D_{nl}(p-p')}\left[ {\frac{{\partial p'_s{B_\textrm{i}}\left( {\beta {\varepsilon _{ip'}}} \right)}}{{\partial {p'_l}}} - \frac{{\partial p_s{B_\textrm{i}}\left( {\beta {\varepsilon _{\textrm{i}p}}} \right)}}{{\partial {p_l}}}} \right]}.
\end{equation}
These expressions show that expansion in the $\sigma$-powers of the operators ${\hat K}_{ab}p_s$ acting \emph{on the even functions} has the form
\begin{align}
{\hat K}_\textrm{ee}^{}p_s&=({\hat K}_\textrm{ee}p_s)^{(0)}+({\hat K}_\textrm{ee}p_s)^{(2)}+O(\sigma^4), &
{\hat K}_\textrm{ei}^{}p_s&=({\hat K}_\textrm{ei}p_s)^{(0)}+({\hat K}_\textrm{ei}p_s)^{(2)}+O(\sigma^4),\nonumber\\
%\begin{equation}
\label{C_K_odd}
{\hat K}_\textrm{ie}^{}p_s&=({\hat K}_\textrm{ie}p_s)^{(1)}+({\hat K}_\textrm{ie}p_s)^{(3)}+O(\sigma^5), &
{\hat K}_\textrm{ii}^{}p_s&=({\hat K}_\textrm{ii}p_s)^{(0)}+({\hat K}_\textrm{ii}p_s)^{(1)}+O(\sigma^3).
%\end{equation}
\end{align}
The explicit expressions for the operators ${\hat K}_{ab}p_s$ are as follows:
\begin{align}
%\[
({\hat K}_\textrm{ee}p_s)^{(0)}g_p &= 2\pi {e^4}m_\textrm{e} L  w_{\textrm{e}p}^{-1}\frac{\partial }{{\partial {p_n}}} \int {\rd^3}{p'} {{w_{\textrm{e}p}}{w_{\textrm{e}p'}}{D_{nl}(p-p')}\left( {\frac{{\partial p'_sg_{p'}}}{{\partial {p'_l}}} - \frac{{\partial p_sg_p}}{{\partial {p_l}}}} \right)}  \nonumber\\
%\]
%\begin{equation}
\label{C_Kee0p}
&\phantom{=}- 2\pi n{z^2}{e^4}m_\textrm{e} L w_{\textrm{e}p}^{-1}\frac{\partial }{\partial {p_n}} w_{\textrm{e}p}D_{nl}(p)\frac{{\partial p_s g_p}}{{\partial {p_l}}},\\
%\end{equation}
%\begin{equation}
\label{C_Kie1}
({\hat K}_\textrm{ie} p_s)^{(1)}g_p &= 2\pi {z^2}{e^4}m_\textrm{e} L w_{\textrm{i}p}^{-1}\frac{\partial }{\partial {p_n}}\int {\rd^3}{p'} w_{\textrm{i}p}w_{\textrm{e}p'}D_{nl}(p')\frac{\partial p'_s g_{p'}}{\partial {p'_l}},\\
%\end{equation}
%\[
({\hat K}_\textrm{ee} p_s)^{(2)}g_p&=- 2\pi {z^2}{e^4}L m_\textrm{e} w_{\textrm{e}p}^{-1}\frac{\partial }{{\partial {p_n}}}\int {\rd^3}{p'} {{w_{\textrm{e}p}}{w_{\textrm{i}p'}}\Big(D_{nl}(p-p'\sigma^2)\Big)^{(2)}\frac{\partial p_s g_p}{\partial {p_l}}}\nonumber\\
%\]
%\begin{equation}
\label{C_Kee2}
&=- 2\pi {z^2}{e^4}m_\textrm{e}^2 n T \sigma^2 L w_{\textrm{e}p}^{-1}\frac{\partial }{{\partial {p_n}}} {{w_{\textrm{e}p}}\frac{\partial^2 D_{nl}(p)}{\partial p_m\partial p_m}\frac{\partial p_s g_p}{\partial {p_l}}},\\
%\end{equation}
%\begin{equation}
\label{C_Kii0p}
({\hat K}_\textrm{ii} p_s)^{(0)}g_p &= 2\pi {z^4}{e^4}m_\textrm{i} L w_{\textrm{i}p}^{-1}\frac{\partial }{{\partial {p_n}}}\int {\rd^3}{p'} {{w_{\textrm{i}p}}{w_{\textrm{i}p'}}{D_{nl}(p-p')}\left( {\frac{{\partial p'_sg_{p'}}}{{\partial {p'_l}}} - \frac{{\partial p_sg_p}}{{\partial {p_l}}}} \right)},\\
%\end{equation}
%\begin{equation}
\label{C_Kei0p}
({\hat K}_\textrm{ei} p_s)^{(0)}g_p &= 2\pi {z^2}{e^4}m_\textrm{e} L w_{\textrm{e}p}^{-1}\frac{\partial }{\partial {p_n}}\int {\rd^3}{p'} {w_{\textrm{e}p}}w_{\textrm{i}p'}D_{nl}(p) \frac{\partial p'_s g_{p'}}{\partial {p'_l}},\\
%\end{equation}
%\[
({\hat K}_\textrm{ii} p_s)^{(1)}g_p &= -2\pi {z^2}{e^4}m_\textrm{e} L w_{\textrm{i}p}^{-1}\frac{\partial }{\partial {p_n}}\int \rd^3 p' w_{\textrm{i}p}w_{\textrm{e}p'}D_{nl}(p') \frac{\partial p_s g_p}{\partial {p_l}}\nonumber\\
%\]
%\begin{equation}
\label{C_Kiip1}
&=-z^3\Lambda m_\textrm{e} T w_{\textrm{i}p}^{-1}\frac{\partial }{\partial {p_n}}w_{\textrm{i}p} \frac{\partial p_s g_p}{\partial {p_l}},\\
%\end{equation}
%\begin{equation}
\label{C_Kiep3}
({\hat K}_\textrm{ie} p_s)^{(3)}g_p &= 2\pi {z^2}{e^4}m_\textrm{e} L w_{\textrm{i}p}^{-1}\frac{\partial }{\partial {p_n}}\int {\rd^3}{p'} w_{\textrm{i}p}w_{\textrm{e}p'}\Big( D_{nl}(p'-p\sigma^2)\Big)^{(2)}\frac{\partial p'_s g_{p'}}{\partial {p'_l}}.
%\end{equation}
\end{align}

The operators $\hat K_{aa}$ take into account the collisions between the particles of the species $a$  both with the particles of the species $a$  and the particles of the species  $b$; the operators $\hat K_{ab}$  take into account only the collisions between the particles of the species $a$ with the particles of the species $b$. The leading order in $\sigma$ of $\hat K_\textrm{ee}$ contains both the electron-electron and electron-ion collisions, but the leading order in $\sigma$ of $\hat K_\textrm{ii}$ contains only the ion-ion collisions~--- this fact is in good accordance with the discussion of the problem by Braginsky \cite{Braginsky1, Braginsky2}.

%\bibliographystyle{cmpj}
%\bibliography{mybibdb}

\clearpage

\ukrainianpart

\title{До релаксаційних явищ у двокомпонентній плазмі}
\author{В.М. Горєв, О.Й. Соколовський, З.Ю. Челбаєвський}
{\address{Дніпропетровський національний університет імені Олеся Гончара, \\ пр. Гагаріна, 72, 49010 Дніпропетровськ, Україна}

\makeukrtitle

\begin{abstract}
\tolerance=3000%
Релаксація температур та швидкостей компонент квазірівноважної
двокомпонентної повністю іонізованої просторово-однорідної плазми
вивчається на основі узагальненого методу Чемпена-Енскога, застосованого
до кінетичного рівняння Ландау. Узагальнення базується на ідеї
функціональної гіпотези Боголюбова з метою врахувати кінетичні моди
системи. Показано, що в наближенні малої різниці швидкостей та
температур компонент системи релаксація дійсно існує (швидкості
релаксації додатні). Доведення базується на аргументах, які придатні
для довільної двокомпонентної системи. Рівняння, які описують
температурну та швидкісну кінетичну моди системи, досліджуються у теорії
збурень за квадратним коренем малого відношення мас електрона та іона.
Рівняння будь-якого порядку в цій теорії збурень розв'язуються методом
розвинення за поліномами Соніна. Отримано корекції до відомих
результатів Ландау, стосовних функцій розподілу компонент плазми та
швидкостей релаксації. Гідродинамічна теорія, яка базується на цих
результатах, повинна враховувати порушення локальної рівноваги при
наявності релаксаційних процесів.

\keywords функція розподілу, метод Чепмена-Енскога, швидкість
релаксації, поліноми Соніна, кінетичне рівняння, кінетичні моди, повністю іонізована плазма

\end{abstract}

\end{document}